\documentclass{aa}
\usepackage{graphicx}
\usepackage[varg]{txfonts}
\usepackage{longtable}
\usepackage{lscape}
\usepackage{amsmath}

\begin{document}

\title{Activity time series of old stars from late F to early K. I. Simulating radial velocity, astrometry, photometry, and chromospheric emission.}

\titlerunning{Impact of stellar activity on radial velocities for old F-G-K stars I.}

\author{N. Meunier \inst{1}, A.-M. Lagrange \inst{1}, T. Boulet \inst{2}, S. Borgniet \inst{3} 
  }
\authorrunning{Meunier et al.}

\institute{
Univ. Grenoble Alpes, CNRS, IPAG, F-38000 Grenoble, France\\
\and
Space sciences, Technologies and Astrophysics Research (STAR) Institute, Universit\'e de Li\`ege,  4000 Li\`ege, Belgium\\
\and
LESIA (UMR 8109), Observatoire de Paris, PSL Research University, CNRS, UMPC, Univ. Paris Diderot, 5 Place Jules Janssen, 92195 Meudon, France \\
\email{nadege.meunier@univ-grenoble-alpes.fr}
     }

\offprints{N. Meunier}

\date{Received 7 December 2018; Accepted 18 March 2019}

\abstract{Solar simulations and observations show that the detection of long-period Earth-like planets is expected to be very difficult with radial velocity techniques in the solar case because of activity. The inhibition of the convective blueshift in active regions (which is then dominating the signal) is expected to decrease toward lower mass stars, which would provide more suitable conditions.}
{In this paper we build synthetic time series to be able to precisely estimate the effects of activity on exoplanet detectability for stars with a wide range of spectral type (F6-K4) and activity levels (old main-sequence stars). }
{We simulated a very large number of realistic time series of radial velocity, chromospheric emission, photometry, and astrometry. We built a coherent grid of stellar parameters that covers a wide range in the (B-V, LogR'$_{\rm HK}$) space based on our current knowledge of stellar activity, to be able to produce these time series. We describe the model and assumptions in detail. }
{We present first results on chromospheric emission. We find the average LogR'$_{\rm HK}$ to correspond well to the target values that are expected from the model, and observe a strong effect of inclination on the average LogR'$_{\rm HK}$ (over time) and its long-term amplitude. }
{This very large set of synthetic time series offers many possibilities for future analysis, for example, for the parameter effect, correction method, and detection limits of exoplanets. }


\maketitle

\section{Introduction}

It is now well recognized that stellar activity strongly affects the detectability of exoplanets. First rough  attempts  to model the amplitude of this effect through radial velocity (RV) have been made with simple models that related simple activity coverage with  jitter \cite[][]{saar97,hatzes02,saar03,wright05}. \cite{desort07} modeled the RV that is caused by single spots. More sophisticated models describing the full behavior of the activity that causes the RV variations are needed, however, to estimate the effect of stellar activity more quantitatively and to test analysis and correction methods. Such models have been made for the Sun \cite[][]{borgniet15} and for a few configurations of other stars \cite[][]{dumusque16,dumusque17}. Other models have been proposed by \cite{herrero16}; they reproduce contributions of spots and plages. \cite{santos15} modeled the contributions of spots alone.

We made a significant step when we modeled the solar RV and photometry using observed solar spots, plages, and network structures \cite[][]{lagrange10b,meunier10a}. This allowed us to show that the inhibition of the convective blueshift in plages dominates the long-term variability, which we validated by reconstructing the solar RV variation from MDI/SOHO (Michelson Doppler Imaging / SOlar and Heliospheric Observatory) dopplergrams \cite[][]{meunier10}. Direct or indirect (Moon, asteroids, Jupiter satellites) observations of the Sun later confirmed these results  \cite[][]{dumusque15,lanza16,haywood16}.
We also studied the effect of activity on future astrometric measurements \cite[][]{lagrange11}, which are important in the context of the current GAIA mission.
  Our second step was to generate similar time series based on randomly generated solar spots and plages, for which we used realistic properties over the solar cycle \cite[][]{borgniet15}: this allowed us to study the effect of inclination, and to open the way to model stars other than the Sun.  This is the objective of the present paper. 
Our approach focuses on using the proper spatio-temporal distribution of spots and plages, and on a physical relationship between spots and plages together with realistic physical properties. This is complementary to other approaches that focus on estimating finer details in contrast variations, for example \cite[e.g.,][]{cegla18}.

In this paper, we therefore extend the solar model described in \cite{borgniet15} to other stars. We propose consistent parameter sets to build RV, photometric, and astrometric time series. We also implement a model to describe the chromospheric emission as a function of time. 
The goal of such simulations is threefold: 1/ to compare our model with observations for these different observables; 2/ to help with the interpretation of these observations, and in particular to understand the degeneracies and biases well, as well as the effect of the various parameters (including "hidden" parameters such as inclination); and 3/ to test correction methods and estimate the effect on exoplanet detectability through various techniques (RV, transits, and astrometry).  
Our objective in building the whole parameter set is to be as consistent as possible in the various choices so that we retain a large amount of the complexity of stellar variability while keeping the parameters to a reasonnable number for this first set: all parameters that correspond to a given time series are compatible with each other. 

The outline of the paper is the following. In Sect.~2 we explain how we adapt the solar parameters to  other stars to generate spots and plages: this section is devoted to the procedures and laws we used to produce lists of spots and plages as well as their properties at each time step. In Section.~3 we describe the  required contrast and RV properties for producing the observables.
We then present the chromospheric emission model in Sect.~4 as well as the calibration that must be made to produce realistic time series. Finally, we compare in Sect.~5 the obtained LogR'$_{\rm HK}$ values with what is expected from the input parameters,  followed by a conclusion and a description of future works in Sect.~6. 


\section{Generation of spots and plages}

\subsection{General principles}

The model we used to produce spots and plages at each time step and to follow their evolution is described in detail in \cite{borgniet15}. We summarize here the main parameter categories: 
\begin{itemize}
\item{{\it \textup{Spatio-temporal distribution of spots and plages:}} butterfly diagram and active longitudes.}
\item{{\it \textup{Long-term variability:}} cycle length, amplitude, and shape.}
\item{{\it \textup{Individual properties:}} size distributions, decay distributions, plage-to-spot size ratio, and plage-to-network decay.}
\item{{\it \textup{Dynamics:}} rotation period, differential rotation, meridional circulation, and diffusion.}
\end{itemize}

Plages are created at the same time as spots, then each type of structure follows its own evolution. Part of the plage decay creates network features. This leads to an entirely coherent model that describes spots, plages, and the network. 
In the following, unless otherwise mentioned (Sect.~2.7), the term plage refers to large plages and network structures, that is, all bright features, in order to simplify the presentation. 
We also recall that \cite{meunier10a} and \cite{borgniet15} considered plage sizes that were obtained from MDI data with a threshold of 100~G, leading to an adjustment of the various contrasts to match the observed solar photometric variations (while associated with the spot distribution). In this paper, we keep size distributions and contrasts that are consistent with that definition. 
The detailed parameters are described in Table~\ref{tab_param} in Appendix E. 
The main differences with the model of \cite{borgniet15} are that we simplified the input spot number (see Sect.~4) and adapted the dispersion that was added to the shape of the reference cycle (see Sect.~2.6.3).

A large number of parameters were involved in our solar simulation. When we adapted these parameters to other stars, we did not have to explore the full space of possible parameters, as some parameters may depend on others. 
For example, the rotation period depends on the spectral type and on the activity level, so that for a given spectral type and activity level, the range of possible periods is limited. 
Empirical laws, sometimes with large dispersion that represents an actual variability between stars, have been established in the literature, allowing us to establish a correspondence between certain variables. 
These relations are sometimes multivariable.  We use these laws in the following to build the parameter sets. 

\begin{figure*}
\includegraphics{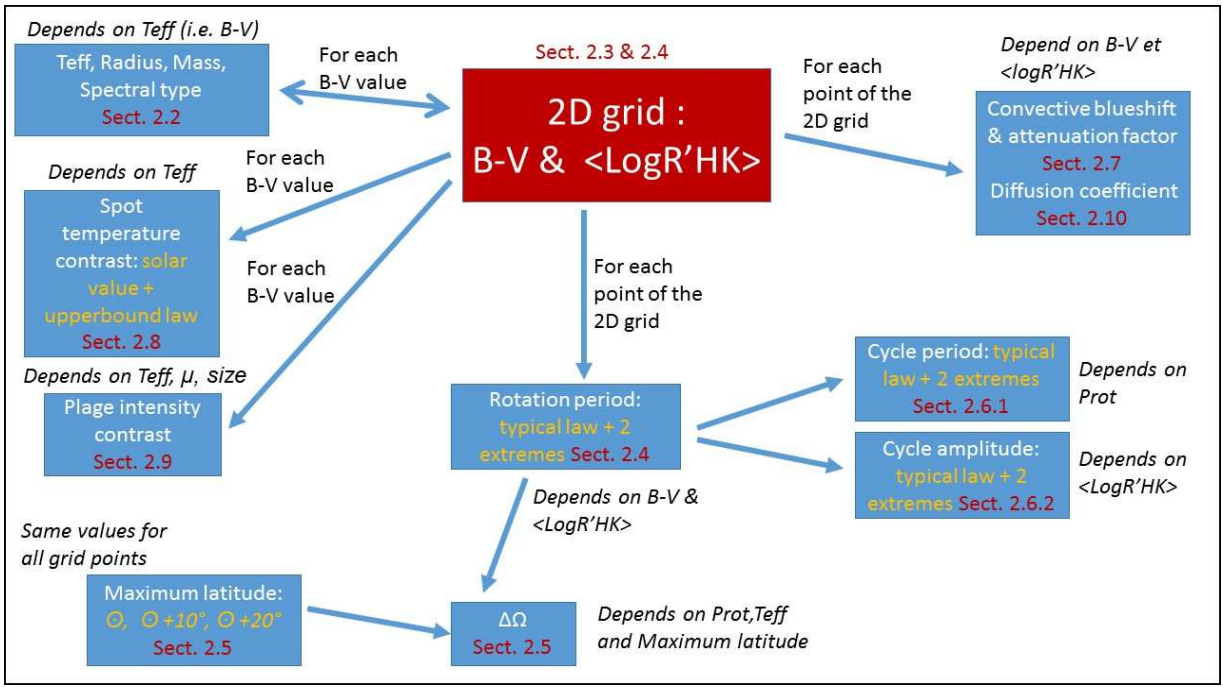}
\caption{
Variable parameters in our grid. Parameters with specific indications in orange depend on B-V and/or LogR'$_{\rm HK}$, and we also consider two or three laws, as indicated. A few parameters depend on specific conditions (plage contrast). The 2D grid in B-V and LogR'$_{\rm HK}$ is detailed in Fig.~3.}
\label{grid}
\end{figure*}

Our objective is to study the effect of important parameters on time series, for example, to establish which types of stars are most suitable for exoplanet detection, or to estimate the performance of correction methods in various conditions. 
Some parameters are not constrained at all for stars other than the Sun, therefore we keep some to the solar values in this work.
The list of parameters that are different from the solar values is described in Fig.~\ref{grid}. A summary of the laws described in the next sections is also given in Table~\ref{tab_law}. 
For five of these parameters, we used two (upper and lower bound) or three laws (median law as well) to cover a range of realistic values because we estimate that the observed variability is real. The remainder of this section is devoted to describing the way in which we derived all these parameters based on our current knowledge of stellar activity. 

\subsection{Fundamental stellar parameters}

\begin{figure}
\includegraphics{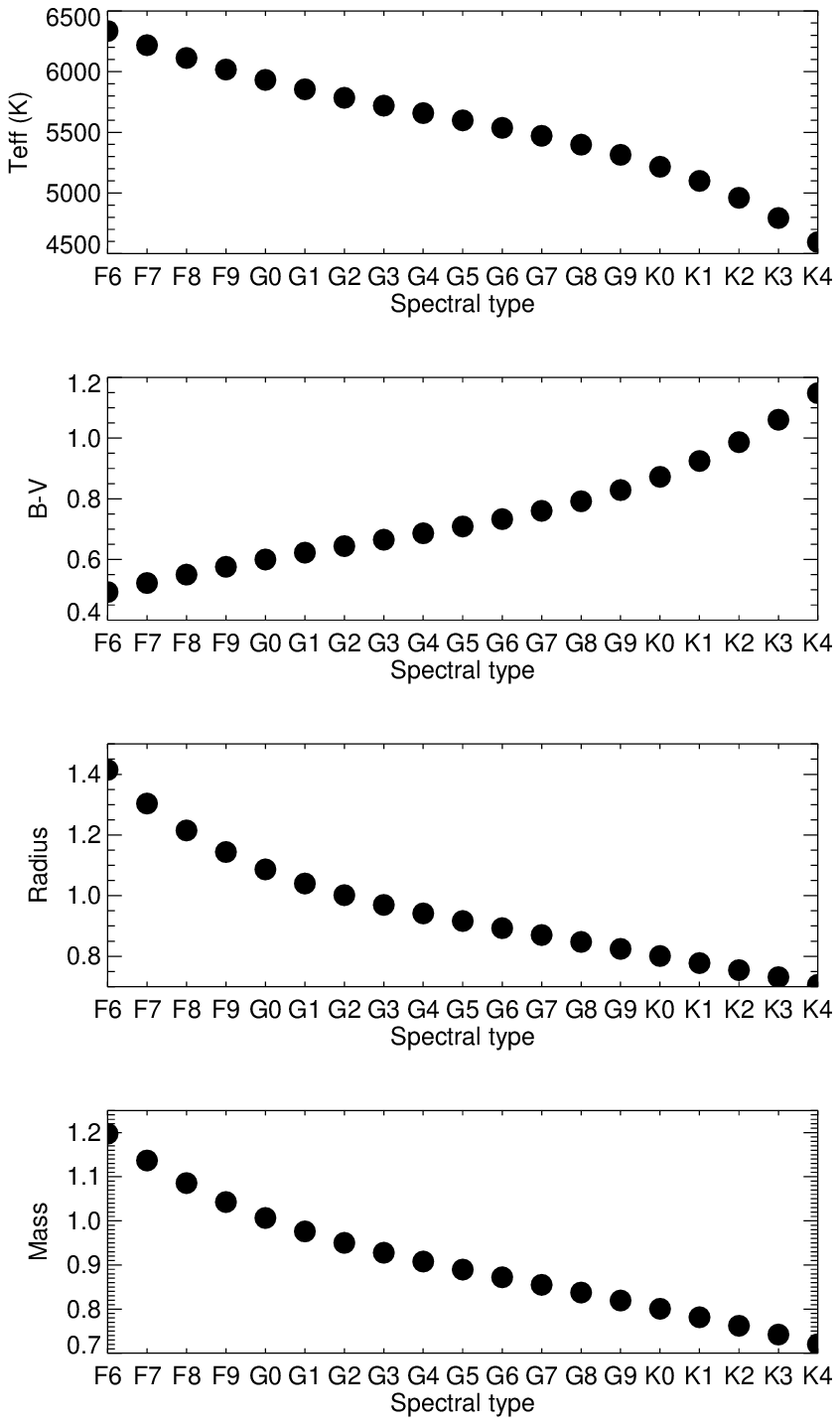}
\caption{
{\it First panel}: T$_{\rm eff}$ vs. spectral type.
{\it Second panel}: B-V vs. spectral type.
{\it Third panel}: Stellar radius (in R$_\odot$) vs. spectral type.
{\it Fourth panel}: Stellar mass (in M$_\odot$) vs. spectral type.
}
\label{fond}
\end{figure}

\begin{table}
\caption{Fundamental parameters}
\label{tab_fond}
\begin{center}
\renewcommand{\footnoterule}{}  
\begin{tabular}{lllll}
\hline
B-V & Spectral & T$_{\rm eff}$ & Radius & Mass \\ 
 & Type    &  (K) &  (R$_{\odot}$) & (M$_{\odot}$) \\ \hline
0.49 & F6 & 6334 & 1.41 & 1.20 \\
0.52 & F7 & 6218 & 1.30 & 1.14 \\
0.55 & F8 & 6112 & 1.22 & 1.09 \\
0.58 & F9 & 6017 & 1.14 & 1.04 \\
0.60 & G0 & 5931 & 1.09 & 1.01 \\
0.62 & G1 & 5854 & 1.04 & 0.98 \\
0.64 & G2 & 5784 & 1.00 & 0.95 \\
0.67 & G3 & 5719 & 0.97 & 0.93 \\
0.69 & G4 & 5658 & 0.94 & 0.91 \\
0.71 & G5 & 5598 & 0.92 & 0.89 \\
0.73 & G6 & 5536 & 0.89 & 0.87 \\
0.76 & G7 & 5470 & 0.87 & 0.85 \\
0.79 & G8 & 5397 & 0.85 & 0.84 \\
0.83 & G9 & 5314 & 0.82 & 0.82 \\
0.87 & K0 & 5215 & 0.80 & 0.80 \\
0.92 & K1 & 5099 & 0.78 & 0.78 \\
0.99 & K2 & 4960 & 0.75 & 0.76 \\
1.06 & K3 & 4793 & 0.73 & 0.74 \\
1.15 & K4 & 4594 & 0.71 & 0.72 \\
\hline
\end{tabular}
\end{center}
\tablefoot{Fundamental parameters for the grid from F6 to K4. }
\end{table}

The spectral type constitutes our first axis in the grid (the second is the average activity level, see the next section). We translate it into B-V values because many empirical laws used in the following are available as a function of B-V in the literature. We consider a wide range of stellar types, F6 to K4, that is, stars whose activity patterns are not very different from that of the Sun.  For example, the convective blueshift, which has a critical effect on the RV amplitudes, has been estimated with a good precision by \cite{meunier17b} for this range of spectral types, but it  is not well constrained beyond this range. 
The four laws we use are illustrated in Fig.~\ref{fond} and the values are listed in Table~\ref{tab_fond}.
\begin{itemize} 
\item{{\it \textup{T}$_{\rm eff}$.} Effective temperatures are derived from the spectral type using a fourth-degree polynomial from the observations of \cite{gray03}. The validity domain is A2--K3, therefore we extrapolate this function over a small range for K4.}
\item{{\it \textup{B-V}.} B-V are derived from T$_{\rm eff}$ (see above) with the law provided by \cite{gray05}.}
\item{{\it \textup{Radius}.} Very many stellar radii have been measured using interferometry \cite[][]{boyajian12,boyajian13}. 
We use Equation 8 from \cite{boyajian12} to relate the radius to T$_{\rm eff}$. This formula  is valid for T$_{\rm eff}$ up to 5500~K, and  we verified that the formula can be extrapolated from the measurements of \cite{boyajian13}: the extrapolation is appropriate up to 6400~K, but with a larger dispersion.}
\item{{\it \textup{Masses}.} 
Stellar masses are derived from the radius in \cite{boyajian12} and \cite{boyajian13} using a third-degree  polynomial fit from their Tables 6 and 3, respectively. 
}
\end{itemize}

\subsection{Average LogR'$_{\rm HK}$ - (B-V) relationship}

The activity level we consider here is the average LogR'$_{\rm HK}$ over time for a given star over time-scales of a few years because a star at a given age does not have a single LogR'$_{\rm HK}$. The average activity level constitutes our second axis. 

\begin{figure}
\includegraphics{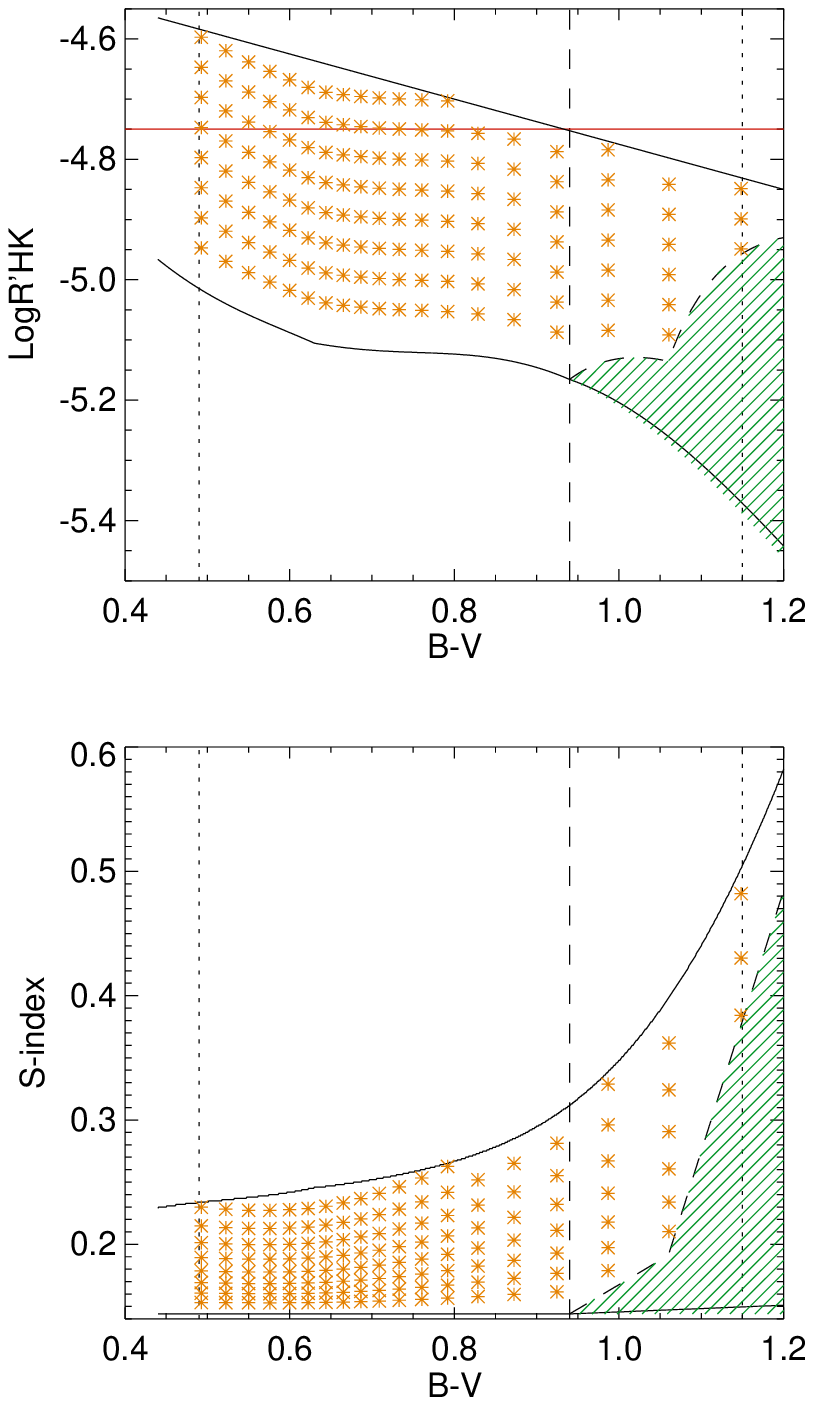}
\caption{{\it Upper panel}: Average LogR'$_{\rm HK}$ versus B-V representing our 2D grid (orange stars). The lower solid line shows the basal flux used in this paper, from \cite[][]{mittag13} and \cite{schroder12} and the upper solid line shows the upper bound for the activity level versus B-V, from \cite{lockwood07}. The dashed line corresponds to the minimum LogR'$_{\rm HK}$ we consider for stars with B-V above 0.94. 
The vertical dotted lines show the range in B-V covered by our simulations, the vertical dashed line vizualises B-V=0.94, and the red horizontal line approximately corresponds to the Vaughan-Preston gap (see text). 
{\it Lower panel}: Same for the S-index versus B-V. 
}
\label{bv_logrphk}
\end{figure}

\subsubsection{Lower limit in LogR'$_{\rm HK}$}

We first estimate the lower limit for the LogR'$_{\rm HK}$. 
Several papers have estimated the average LogR'$_{\rm HK}$ versus B-V for large samples of stars and dates \cite[][]{henry96,gray03,gray06,jenkins08,isaacson10,jenkins11,arriagada11,schroder12,mittag13}. We first consider the B-V range from 0.45 to 0.94. \cite{mittag13} obtained a flat minimum S-index in this domain. This limit is not entirely strict because occasionally, a few stars lie below it, but this lowest flux derived from \cite{mittag13} is consistent with previous publications. Therefore we consider their value (corresponding to a S-index of 0.144) in the following to be the lower limit for activity in this B-V domain.
This S-index can then be converted into a LogR'$_{\rm HK}$ value using the commonly used formula from \cite{noyes84}, as we do here. 

Finally, we consider B-V above 0.94. A strong increase of the lower limit in activity \cite[][]{isaacson10,mittag13} corresponds to stars with a significant degree of activity, implying that low-activity stars are not observed in this B-V range. We have checked the HARPS spectra of stars close to this apparent limit, and they indeed show strong calcium emission. This lower limit can therefore be used to identify where stars are located in the 2D space (B-V, LogR'$_{\rm HK}$). 
 In conclusion, the considered LogR'$_{\rm HK}$ values lie above the dashed line in Fig.~\ref{bv_logrphk} (solid for B-V below 0.94 as the two coincide in that domain).




\subsubsection{Upper limit in LogR'$_{\rm HK}$}

We now consider the upper limit in LogR'$_{\rm HK}$.  
A first simple choice would be to consider a threshold from the Vaughan-Preston gap in the usually bimodal distribution of LogR'$_{\rm HK}$ values. Depending on the publication, the position of the gap ranges from -4.80 \cite[][]{noyes84,jenkins11} to -4.6 or -4.7 \cite[][]{mamajek08} with intermediate values \cite[][]{wright04,jenkins08,henry96,gray03,gray06}. However, our purpose is to model stars with properties similar to solar properties in terms of plage-to-spot ratio, for example, at least given our current knowledge. We therefore used the results from \cite{lockwood07}, which show this type of correlation versus B-V and logR'$_{\rm HK}$. The interface between the spot-dominated regime (younger stars) and the plage-dominated regime (the older stars we are interested in) varies with B-V, and is about -4.5 for the most massive and -4.85 for the less massive stars. We use this as an upper limit for our LogR'$_{\rm HK}$ values (shown as the upper solid line in Fig.~\ref{bv_logrphk}). 

We could also have derived this upper limit from age isochrones \cite[][]{mamajek08}, but because we are more interested in the plage-to-spot ratio in our input parameters, this choice would be less pertinent. The age range covered by our simulations may therefore vary with B-V (see Sect.~2.4).

\subsubsection{LogR'$_{\rm HK}$ values between the lower and upper limits}

Stars are observed with LogR'$_{\rm HK}$ between the lower and upper limits that we defined in the previous sections. The distribution of stars within that domain is not necessarily homogeneous, but this was not taken into account when we built the grid. 

We considered LogR'$_{\rm HK}$ values higher than the lower level by 0.07~dex and then with a step of 0.05~dex up to the upper bound. Theses values are shown as orange stars in Fig.~\ref{bv_logrphk}. They lead to 141 points in 2D space and correspond to the average LogR'$_{\rm HK}$ over time. For each of these positions, parameters were defined according to Fig.~\ref{grid} and several time series were built. These parameters are described in detail in the remainder of this section.

\subsection{Rotation period versus B-V and LogR'$_{\rm HK}$}

\begin{figure}
\includegraphics{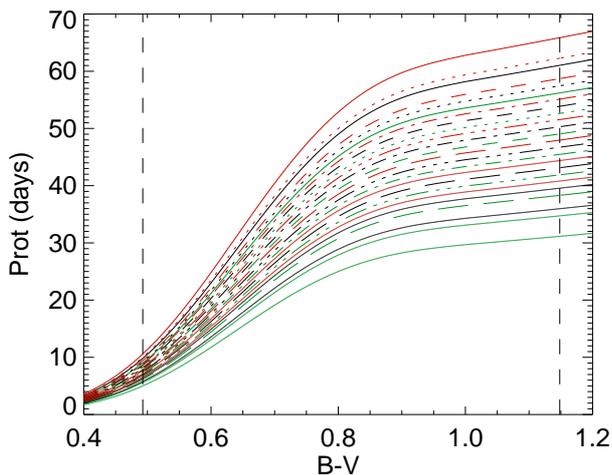}
\caption{
Chosen rotation periods vs. B-V for eight different logR'$_{\rm HK}$ values between -5.1 and -4.75 (from top to bottom) from \cite{mamajek08}. The median law is shown in black, and extreme laws are shown in green and red. 
}
\label{prot}
\end{figure}

 In this section we wish to determine which rotation rate (or range of rotation rates) to use to simulate a star of a given spectral type and average activity level. 
Several estimates of the rotation period as a function of the average activity level have been published using large samples of stars, either directly or through an estimate of the Rossby number. A comparison of these different laws is provided in Appendix A. We have then chosen to use the law relating the Rossby number and the average LogR'$_{\rm HK}$ from \cite{mamajek08}, with the estimated turnover time from \cite{noyes84} to relate the rotation period and the Rossby number. 
 Long periods are difficult to estimate, and samples are usually biased toward short periods. Laws are therefore uncertain for long periods, which correspond to our lower mass stars. We have then taken into account the observed dispersion around the law provided by \cite{mamajek08}, which we estimated from their data to be of about $\pm$0.2 in Rossby number: from these upper and lower bound laws we derived the rotation period versus B-V for each LogR'$_{\rm HK}$ , which we show in Fig. 4 as red and green curves, respectively.  
 Although we took the observed dispersion into account, we might still underestimate the longest rotation periods: if this were the case, the effect on the final RV or photometric jitter is expected to be very small. However, it is expected to affect the frequency analysis in two ways: the power due to rotation will naturally be localized at a different period, and this may affect the morphology of the curves because the ratio between the rotation rate and the typical lifetime of the magnetic features will be different. The fact that simulations are always made for three different rotation rates will help to analyze the effect of our assumptions in future works. 

Age is not a parameter in our simulation. However, we know that there is a relationship between rotation, activity level, and age \cite[e.g.,][]{wilson63,skumanich72}. For instance, when we use the laws of \cite{mamajek08} 
for the most massive stars in our simulations, our range in LogR'$_{\rm HK}$ corresponds to ages between 0.5 and 3 Gyr. Lower mass stars in our simulations correspond to older stars, typically between 4 and more than 10 Gyr,  depending on their average activity level. 


\subsection{Differential rotation and latitude coverage}

\begin{figure}
\includegraphics{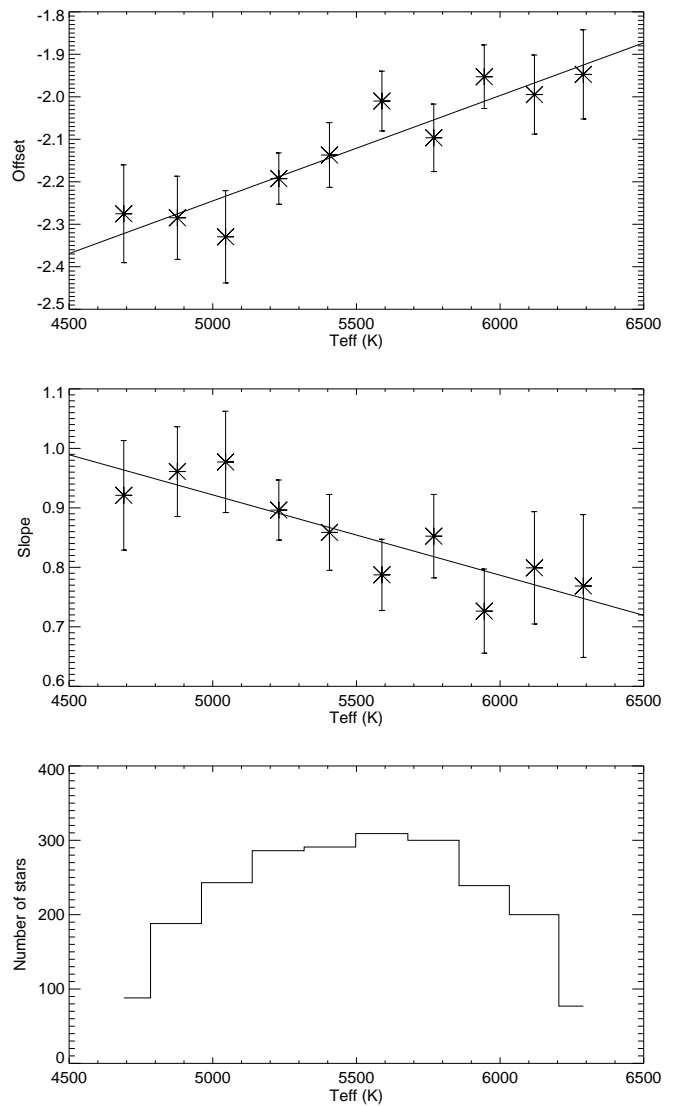}
\caption{
{\it Upper panel}: Coefficient $p_0$ vs. T$_{\rm eff}$ from the fit of Log($\alpha$)=$p_0$+$p_1 \times$Log(P$_{\rm rot}$) for stars with log(g) between 3.94 and 4.94. Computations are made from the data published by \cite{reinhold15}.
{\it Middle panel}: Same for $p_1$.
{\it Lower panel}: Number of stars in each T$_{\rm eff}$ bin.
}
\label{alpha2}
\end{figure}

The implementation of differential rotation is strongly related to the latitudinal extension over which magnetic activity is present because measurements of the differential rotation, based on the presence on active structures, only provide the differential rotation over that range in latitude and not the value corresponding to a full range of 0-90$^{\circ}$. In this section we therefore discuss these two parameters together. 

\subsubsection{Differential rotation versus temperature}

 To derive a practical relation using our other input parameters, we used the differential rotation measured by \cite{reinhold15} from Kepler data for a very large sample of stars. As for the rotation period, we should keep in mind that observations are biased toward active stars, that is, fast rotators. Our objective is to define a law $\Omega (\theta),$ where $\Omega$ is the rotation rate and $\theta$ is the latitude that can be used in our simulations. A parameter $\alpha$ is commonly defined from the minimum and maximum rotation periods given in \cite{reinhold15}, $P_{\rm min}$ and $P_{\rm max}$ ,
\begin{equation}
\alpha = \frac{P_{\rm max}-P_{\rm min}}{P_{\rm max}}
,\end{equation}
which is a relative differential rotation. $\alpha$ is then available as a function of T$_{\rm eff}$. 
We note that the $\Omega (\theta)$ function for the Sun is usually described with three parameters \cite[e.g.,][]{snodgrass90} as $\Omega(\theta) = \Omega_0+\Omega_1 sin^2(\theta)+\Omega_2 sin^4(\theta)$. Because the differential rotation for stars is much less well defined, we used only the first two coefficients in the following ($\Omega_2$=0). When the following results are compared with solar differential rotation, caution is therefore advised. 

\begin{figure}
\includegraphics{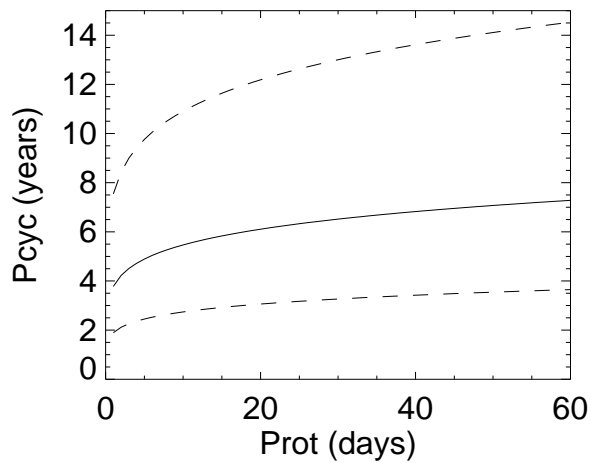}
\caption{
P$_{\rm cyc}$ (in years) vs. P$_{\rm rot}$ used in our grid (upper bound as solid line, lower bound and intermediate range as dashed lines). 
}
\label{pcyc}
\end{figure}

We considered the stars in the sample from \cite{reinhold15} with log(g) between -3.94 and -4.94 as in some previous analyses of solar type stars \cite[][]{daschagas16}. For ten bins in T$_{\rm eff}$ in our range in temperature, we performed a linear fit between Log($\alpha$) and Log(P$_{\rm rot}$), where P$_{\rm rot}$ is defined as
\begin{equation}
P_{\rm rot}=\frac{P_{\rm min}+P_{\rm max}}{2} 
.\end{equation}

The two coefficients of these linear fits, $p_0$ and $p_1$, are shown in Fig.~\ref{alpha2} as a function of T$_{\rm eff}$. We then modeled $p_0$ and $p_1$ as a linear function of T$_{\rm eff}$, which gives 
\begin{equation}
p_0=-3.485+2.478 10^{-4} \times T_{\rm eff}
\end{equation}
and
\begin{equation}
p_1=1.597-1.35 10^{-4}\times T_{\rm eff}
.\end{equation}

For each T$_{\rm eff}$ and P$_{\rm rot}$ (previous section) in our grid, we can then derive $\alpha$. We discuss differential rotation in more detail in Appendix B. 

 The effect of our choice of $\Omega(\theta)$ is not expected to be critical for our simulations.  When differential rotation is present, periodograms of  the time series are thought to exhibit multiple peaks around the rotation period, which complicates estimating the rotation period, for example (in addition to the limited lifetimes of structures). The precise choice of $\Omega(\theta)$  therefore mostly affects the complexity of the peak structures in the periodograms, but it does not affect the signal amplitude, for example.

\subsubsection{Maximum latitude}

We assumed 1/ that structures are always present at low latitude at the end of the cycle \cite[when the maximum latitude is higher, there might be no activity close to the equator either, as shown, e.g., by][but these effects are not full understood so far]{isik11}, and 2/ that the latitude coverage is directly related to the maximum latitude of the butterfly diagram (in the case of the Sun, we used an average latitude at the beginning of the cycle of 22$^{\circ}$, with a possible extension of activity to 42$^{\circ}$), hereafter $\theta_{\rm max}$. How $\theta_{\rm max}$ varies with T$_{\rm eff}$ is not constrained. For lower mass stars, where the convective zone is thicker, we expect higher values of $\theta_{\rm max}$  \cite[e.g.,][]{isik11}. On the other hand, for a shorter P$_{\rm rot}$ (in our case, for higher mass stars), we also expect larger $\theta_{\rm max}$  \cite[e.g.,][]{schussler92}. Because these two effects compete with each other, we do not know the proper trend from observations or numerical simulations. Simulations such as the one made by \cite{isik11} are expected lead to some results in the future, but so far, the coverage in parameters is too sparse to conclude. As for the observation, the analysis of the data results of \cite{reinhold15} does not allow us to conclude either. Recent results using planetary transit across spot at the stellar surface allowed determining the latitudinal distribution for a specific star over a short period of time, see \cite{morris17}, but the statistics is not yet sufficient to ascertain any trend.

 Because the coverage in latitude of magnetic structures in not well constrained by either observations or observations,  we considered in our simulations  three possible levels for $\theta_{\rm max}$, remembering that we do not know how other stars differ from the Sun in that respect: the solar value itself $\theta_{\rm max,\odot}$, $\theta_{\rm max,\odot}$+10$^{\circ}$ , and $\theta_{\rm max,\odot}$+20$^{\circ}$. For each of these values of $\theta_{\rm max}$, P$_{\rm rot}$ (derived from the previous section and as defined in equation 2) and $\alpha$ (derived as described above and defined in equation 1) lead to $\Omega_0$ and $\Omega_1$, assuming that $P_{\rm max}$ corresponds to $\theta_{\rm max}$ and $P_{\rm min}$ corresponds to the equator. This allows us to estimate the effect of these parameters on the time series.

%
%
%

\subsubsection{Antisolar rotation}

Numerical simulations have indicated that some stars probably present antisolar differential rotation (rotation is slower at the equator than at the poles). This would occur for stars with large Rossby number, that is, long P$_{\rm rot}$ and high masses \cite[e.g.,][]{brun17}. 
This is very difficult to observe from light curves \cite[][]{santos17}, however, although there have been a few indications that this could be present: \cite{reinhold15b} made tests on synthetic time series, applied their method to a small sample of 50 Kepler stars, and estimated that there was a possibility that the rotation in 10-20\% of the stars is antisolar. When we compare our  parameter grid in the (P$_{\rm rot}$, stellar mass) space with the results from the magnetohydrodynamics (MHD) simulation of \cite{brun17}, we find that fewer than 6\% of our simulation stars (all are less massive than the Sun and are very quiet) may have such an antisolar differential rotation, although the threshold between the solar and antisolar regimes is not well defined.  
It is therefore still very uncertain, especially for stars with our parameter range, and probably does not concern many stars. Because this effect would not significantly affect our results, we consider only solar differential rotation here.

\subsection{Cycle properties}

The cycles of the stars we simulated are similar to the solar cycle in shape, although the amplitudes and ratio between maximum and minimum may be different. Stars with no variability will not be reproduced adequately, although the simulation with a very low cycle amplitude will present some similarities with such stars. 

How many stars have a cycle is subject of debate. However, the existence of long-term variations is crucial for us here because these variations are critical for studying the effect on exoplanet detectability. \cite{baliunas98} analyzed Mount Wilson data and found that 15\% of the stars had a constant activity level, 25\% had a variability without any obvious periodicity (they did not show any smooth cycle like the Sun, but rather some erratic variations, and they correspond to young fast-rotating stars), and 60\% had solar-like cycles.
 \cite{lovis11b} were unable to find any period (defined as the period derived from the fit of a sine function on the data, even if a single cycle was observed) for 66\% of  a large sample of variable stars observed with HARPS, which is likely due to the sampling: Stars without an identified period are very strongly biased toward a very poor sampling compared to the list of stars with an identified "cycle" ("cycle" here means that a proper fit with a sinusoidal was possible, not that it was repeating itself). When this bias is taken into account, the percentage of stars without a long-term variation similar to a "cycle" decrease to ony 15-20\%. 
More recently, results obtained from the analysis of the long-term Mount Wilson survey together with the Lowell survey \cite[][]{hall19} show that 40\% of the stars may have a relatively flat chromospheric emission over decades, although some of them show high chromospheric emission. 

 The statistics of the various stellar categories is therefore still uncertain. In practice, the lower cycle amplitude in our grid will allow us to cover almost no variability reasonably well, at least at low average activity level, because the ratio between the number of spots at cycle maximum and at cycle minimum will be able to reach values close to 1 in some cases. We do not attempt here to add more complexity to our time series, as this would represent additional parameters that are not at all constrained (e.g., what does the butterfly diagram look like when two cycle periods are present?), but future work will have to consider these configurations more precisely. 
Therefore our simulations are quite representative of stars that have some significant variability, except for the more complex stars, as well as of stars with very low variability. 



\subsubsection{Cycle period}

 We compared the cycle period versus rotation rate from various sources (\cite{baliunas96,saar99,bohm07,olah09,olah16,suarez16}) for the so-called inactive branch when relevant. Stars on this branch are old stars similar to the Sun as in our model, and not young active stars.  Except for \cite{bohm07}, which lies apart, these stars provide a coherent picture. The slope of Log(P$_{\rm cyc}$/P$_{\rm rot}$) versus Log(1/P$_{\rm rot}$) is in the range 0.74--1.09, and we consider here an average between these different sources, that is, a coefficient of 0.84: this gives the cycle period we show in  Fig.~\ref{pcyc}. The curve is relatively flat. However, the dispersion in the observations is likely to be real, and the dashed lines shows the two extreme laws that we also considered to account for the observed variability. We therefore explored a wide range of cycle periods.

\subsubsection{Cycle amplitude}


\begin{figure}
\includegraphics{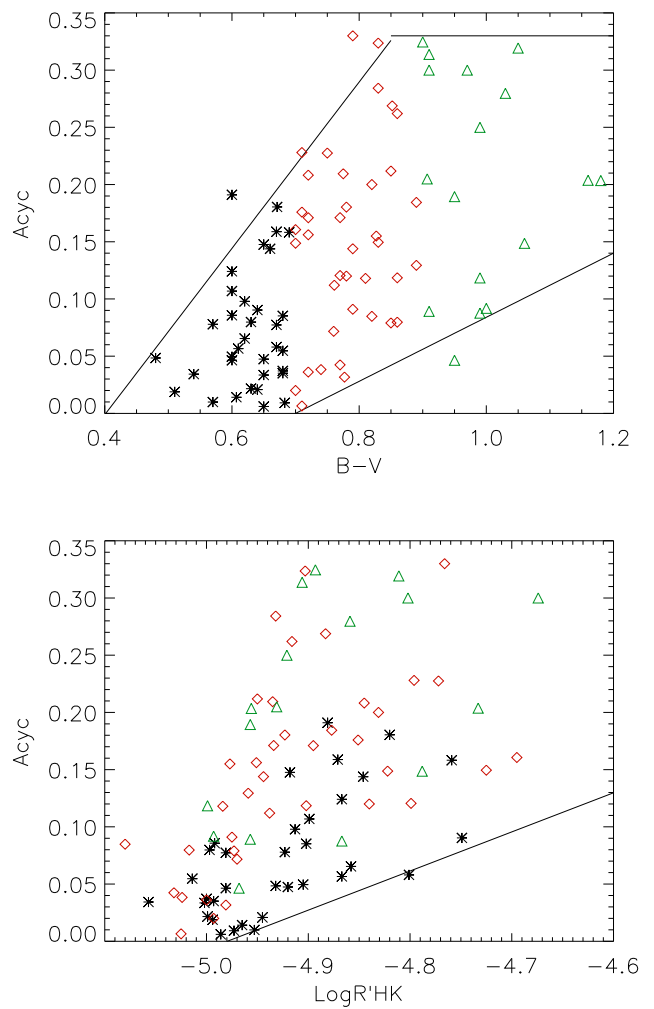}
\caption{
{\it Upper panel}: Half full amplitude of stellar cycles vs. B-V, derived from \cite{lovis11b} after revision of the largest amplitudes (see text) for different types of stars: B-V $<$ 0.7 (black stars), 0.7$<$ B-V$<$ 0.9 (red squares), and B-V $>$0.9 (green triangles). The black lines correspond to the lower and upper boundaries that were taken into account in building the grid. 
{\it Upper panel}: Same vs. average LogR'$_{\rm HK}$. The solid line represents the lower limit that was taken into account in building the grid.
}
\label{acyclovis}
\end{figure}


Several studies produced amplitudes for the cycle period, especially as a function of LogR'$_{\rm HK}$. We have compared the laws from various sources: \cite{radick98}, \cite{saar02}, \cite{lockwood07}, \cite{hall09}, and \cite{lovis11b}. When the observed dispersions are taken into account, the agreement between them is good overall, except for the existence of very large amplitudes in \cite{lovis11b}; we discuss this below. The trend for large amplitudes for larger LogR'$_{\rm HK}$ is globally weak. The values obtained by \cite{hall09} seem to be slightly lower than in other studies. It is also important to note that in general, these samples contain very few quiet stars with LogR'$_{\rm HK}$ below -5.0, so that the amplitudes in this domain are not well constrained (and are also most likely to be affected by noise).

Because the very large upper bound derived by \cite{lovis11b} is very puzzling, we verified the temporal variability of all stars whose A$_{\rm cyc}$ (half-amplitude in R'$_{\rm HK}$x10$^5$)  was larger than 0.3 in their sample. To do this, we used HARPS archive data. We find that with extended observations since 2011, all of them fall below 0.33, which agrees very well with the other publications.

Fig.~\ref{acyclovis} shows A$_{\rm cyc}$ from \cite{lovis11b} versus B-V and average LogR'$_{\rm HK}$. Stars of different spectral types have different A$_{\rm cyc}$. We used the boundaries indicated in the figure to derived three laws: an upper value, a lower value, and an intermediate between the two for each point of the grid in (B-V, LogR'$_{\rm HK}$). We chose a minimum A$_{\rm cyc}$ of 0.005: this corresponds to stars with very low variability. 

\subsubsection{Cycle shape}

\begin{figure}
\includegraphics{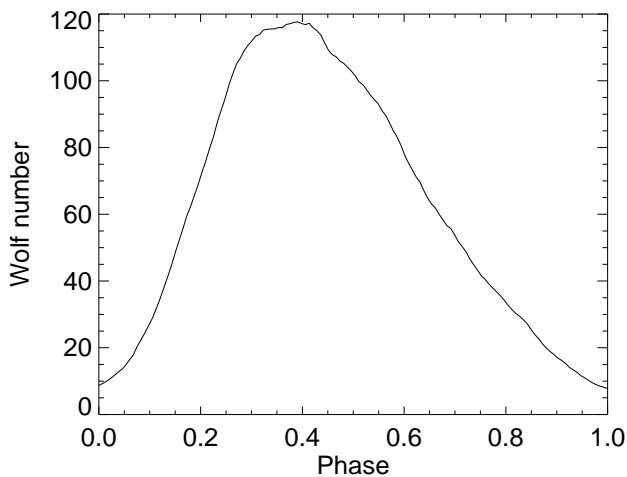}
\caption{Wolf number vs. phase smoothed over the last solar cycle that was used as a reference.}
\label{formcyc}
\end{figure}

As discussed at the beginning of this section, we considered cycles similar to the solar cycles. The chosen shape of the cycle is the shape of the last solar cycle, as shown in Fig.~\ref{formcyc}. The ratio between maximum and minimum can be different, however. At each time step, some random variability was added to that curve (25\% of the amplitude at that time step) to represent the stochastic variability that can be introduced by the dynamo in terms of flux emergence. This amplitude is somewhat arbitrary, but it gives a final realistic dispersion for the Sun.  The input parameters, in addition to this shape and dispersion (constant), are therefore the minimum and maximum level in LogR'$_{\rm HK}$, which must  correspond to the average LogR'$_{\rm HK}$ we wish to obtain.

\subsection{Small-scale convection level and convective blueshift}

\begin{figure}
\includegraphics{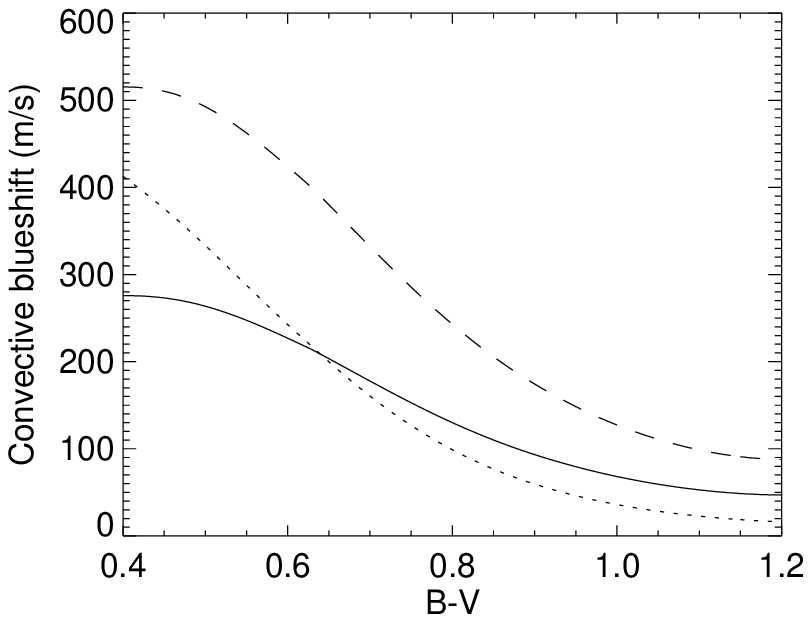}
\caption{Convective blueshift derived for the basal logR'$_{\rm HK}$ (dashed line), adapted from \cite{meunier17b}, and scaled with the solar values derived in \cite{meunier17}. The solid line shows the local convective blueshift we used in our model, after correction for projection effects for a constant attenuation of the convective blueshift in plages of 0.38 (see text). The dotted line shows the same parameter when the trend of the attenuation factor vs. T$_{\rm eff}$ is taken into account.
}
\label{convbl}
\end{figure}


The inhibition of the convective blueshift in plages is an important contribution to the final RV. This parameter has no effect on the other observables. We used the study of the activity effect on convective blueshift based on a large sample of F-G-K stars that was presented in \cite{meunier17b}. We found that the convective blueshift depended not only on B-V, but also on LogR'$_{\rm HK}$. For several B-V bins, we extrapolated these convective blueshifts to the basal LogR'$_{\rm HK}$ (i.e., the level of convective blueshift we would have if no activity were present). This gives the dashed line in Fig.~\ref{convbl}. The convective blueshift was estimated as in \cite{meunier17}, that is, it is based on the Sun from \cite{reiners16}, with 355~m/s for the solar convective blueshift. 

After we derived the convective blueshift, we applied an attenuation factor (which provides the amplitude of the RV in plages and network structures) and a correction factor for projection effects (considering effects perpendicular to the surface, as in our previous work). 
In previous work, we used an attenuation factor of two-thirds based on \cite{BS90}, but also a smaller solar convective blueshift. Because our amplitude led to good results when we compared out results with a solar reconstruction of the long-term RV variation \cite[][]{meunier10}, we would need to use a smaller attenuation factor (0.38) to obtain the same results, given our new convective blueshift: this is what was used in our simulations, which gives the local $\Delta V$ applied to each structure as a function of B-V, shown as the solid line. 

Another result obtained by \cite{meunier17b} is a possible trend versus T$_{\rm eff}$ for the attenuation factor, which would imply a correction factor of -2.077+5.324 10$^{-4}$ T$_{\rm eff}$. The effect of this trend is shown in Fig.~\ref{convbl} as the dotted line. However, this trend is poorly constrained below 5300 K, and we therefore chose to make our simulation without this factor. The effect on the resulting RV can be estimated during analysis, as this correction can be applied afterward to the time series.

Finally, the convective blueshift is higher in larger structures. We implemented the dependence between the velocity and the size derived in \cite{meunier10}. There is typically a ratio of 6 between the largest and smallest structures. 

\subsection{Spot temperature}

\begin{figure}
\includegraphics{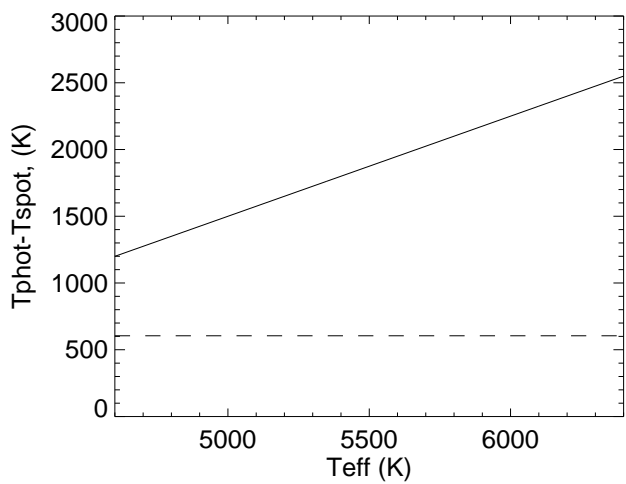}
\caption{Upper and lower bounds for the difference between spot and photosphere temperature vs. T$_{\rm eff}$ from \cite{berd05}. The lower bound is the solar value derived in \cite{borgniet15}.
}
\label{tspot}
\end{figure}

The temperature of stellar spots remains poorly constrained because it is very difficult to measure (and strongly degenerated with spot size in photometric light curves). We used the results of  \cite{berd05}, which show a trend with T$_{\rm eff}$ (lower spot contrast for lower stellar T$_{\rm eff}$), and a large dispersion because stars (including the Sun) exist at a much lower temperature contrast. The trend and order of magnitude have recently been confirmed in numerical spot simulations \cite[][]{panja19}. We therefore used two laws that represent two extreme configurations, assuming that stars have temperatures within that range. This is shown in Fig.~\ref{tspot}: the lower boundary law corresponds to the solar contrast we used in \cite{borgniet15}, and  the upper boundary is derived from \cite{berd05}. The computations were made at 6000\AA, as in \cite{borgniet15}, and the contrasts were adjusted to correspond to the bolometric photometric variability \cite[see][for details]{borgniet15}.


\subsection{Plage contrasts}

Contrasts of stellar plages are poorly constrained as well. In our previous work for the Sun \cite[][]{meunier10a}, we used a law that described a temperature contrast versus $\mu$ (cosine of the angle between the line of sight and the local vertical at the solar surface) similar to the law described in \cite{unruh99}. We then adjusted this slightly (together with the contrast of the spot temperature that we described in the previous section) to fit the observed solar irradiance. \cite{borgniet15} used a  description in terms of intensity contrast (plage intensity divided by quiet-Sun intensity, minus one), which was described as a second-degree polynomial in $\mu$. A similar adjustment was made to fit the photometry (which was necessary because we used a slightly different center-to-limb darkening function). 

In the present paper, we not only need to describe the contrast as a function of $\mu$ for the Sun, but also for other spectral types. We used the results from the MHD simulations performed with the MURAM code by C. Norris \cite[][]{norris16,norris17,norris18}. She provided coefficients describing the plage intensity for different magnetic field levels (0~G, 100~G, and 500~G) versus $\mu$ (in the 0.2-1 range, not available for $\mu$ below 0.2) for G2, K0, and M0 stars as described in \cite{norris18}. These intensities where computed for the HARPS wavelength range. The contrasts take slightly different shapes depending on the parameters, but the global trend is a higher contrast for stronger magnetic fields and higher T$_{\rm eff}$. 

Different functions can be used to describe the intensity variations versus $\mu$ \cite[e.g.,][]{yeo13}, and they provide different values in the 0-0.2 $\mu$ range: however, their effect on the final RV is very low (lower than 0.1\%), therefore we use a quadratic form in $\mu$ in the following. We then compute the contrasts as quadratic functions of $\mu$ for these three spectral types and two magnetic field levels and interpolate (or extrapolate for stars in the F6-G1 range) for other spectral types and different magnetic fields. 

Our simulations provide sizes. We therefore established a law relating the size ($A$) and the magnetic field flux ($B$) for the plages and magnetic features we are interested using MDI/SOHO \cite[][]{Smdi95} magnetograms that cover a full solar cycle, which gives
\begin{equation}
Log(B) = 2.1134+0.1355*Log(A) 
,\end{equation}
where B is in G and A in ppm of the hemisphere. For each structure in the simulation, we therefore computed its associated magnetic field according to this law, to be able to interpolate (all values are between 100 and 500~G). 

When we apply this procedure to a G2 star with an activity level similar to that of the Sun, we find that the contrasts are slightly higher on average than those of \cite{borgniet15}, they are higher by a factor 1.5. We therefore divided the contrasts by this value for all stars for consistency with our definition of structures sizes and the good agreement with the solar irradiance variability (and the corresponding definition of spot size and spot contrast).

\subsection{Other parameters}

\begin{itemize}
\item{ {\it \textup{Spatio-temporal distribution}}. Latitude and longitude distributions as well as north-south asymmetry and active longitude parameters were kept to the solar values from \cite{borgniet15}. The latitude at the beginning of the cycle (related to the maximum possible latitude) was also tested with different values (see Sect.~2.5.2). Migration was considered to be equatorward as for the Sun, although there have been indications that poleward migration could exist \cite[][]{messina03,moss11}, but this is not well constrained.
}
\item{ {\it \textup{Large-scale dynamics}}. The differential rotation discussed in Sect.~2.5 that we adapted to each grid point is described with only two  coefficients (instead of three for the Sun). The same law was used for all structures (spots, plages, and network). The meridional circulation was kept to the solar value used in \cite{borgniet15} based on \cite{komm93}, and was also the same for all structures.
}
\item{ {\it \textup{Spot properties}}. 
The distributions of spot size and decay rate were kept to the solar values used in \cite{borgniet15}, which were adapted from \cite{martinez93}, \cite{baumann05}, \cite{lagrange10b}, and \cite{meunier10a}. 
}
\item{ {\it \textup{Faculae properties}}. The faculae properties were similar to those used in \cite{borgniet15}, which included the ratio distribution between plage and spot sizes, and decay rates (here we considered the plages that were produced each time we generated a spot). 
}
\item{ {\it \textup{Network properties}}. Network properties were similar to those used in
 \cite{borgniet15}, which include the diffusion coeficient from \cite{schrijver01}, the fraction of plage flux that was used to build the network, and the decay rate.  The diffusion coefficient was scaled with the amplitude of the convective blueshift, as discussed in Sect.~2.7. 
}
\end{itemize}

\section{From structures to observables}

\subsection{Filling factor and photometric and radial velocity time series}

Because of the huge number of simulations, it is not possible to compute the observables as done in our previous work \cite[][]{lagrange10b,meunier10a,borgniet15}, in which we computed maps, then spectra, and finally RVs (and other observables) in the same way as for stellar observations. We therefore simplified the computations as follows: we directly summed the contribution from each structure to the RV, photometry, and astrometry. The filling factors were also computed. This implies that we assumed that the structures are point-like, which means that we did not need to check for superimpositions. This is different to what was done in \cite{borgniet15}.  This also means that we neglected any geometrical effect that would be due to a large area covered by a structure, which is a good assumption because we modeled relatively quiet stars with moderate structure sizes. The assumption is very good for spots because the maximum size of a spot would correspond to a radius of about 3$^{\circ}$ (which would be a very rare case; typical spots are much smaller, see Table~\ref{tab_param}). Plages may cover a larger area, which is expected to add a second-order distortion to an ideal time series: in most cases, the extreme sections of a structure (east and west) are expected to produce a signal that is very similar to the central part of the structures on average; in addition, we considered at each time step an irregular decay of the structures that has a random amplitude, so that the distortion produced by the assumption would not be identifiable given this other source of irregularity. The same structures were used for all stellar inclinations.  The formulae are provided in Appendix C.

\begin{table}
\caption{Variable parameters}
\label{tab_param}
\begin{center}
\renewcommand{\footnoterule}{}  
\begin{tabular}{lll}
Parameters & Main grid & Calibration grid\\ \hline
LogR'$_{\rm HK}$ vs B-V & 141 values & 19 B-V  \\
P$_{\rm rot}$ & 3 laws & 1 law \\
P$_{\rm cyc}$ & 3 laws & - \\
A$_{\rm cyc}$ & 3 laws & 1 activity level \\
$\theta_{\rm max}$ (\&  $\Delta \Omega$) & 3 $\theta$ & 3 $\theta$  \\
TOTAL & 11421 & 57 \\ \hline
{\it Tspot} & {\it 2 laws} & - \\
{\it Inclinations} & 10 & 10 \\
\hline
\end{tabular}
\end{center}
\tablefoot{Number of values or laws determining the number of simulations. The spot temperature is related to the observable and not to the generation of structures, therefore the two values considered are used for the same list of structures. The first column (main grid) corresponds to the parameter sets described in Sects.~2 and 3. The second column corresponds to the simulations dedicated to the calibrations described in Sect.~4. }
\end{table}

We recall that we chose to study different laws for several of our parameters (as summarized in Fig.~\ref{grid}), so that several time series of spots, plages, and network features were produced for each of the 141 points of the 2D grid in (B-V, LogR'$_{\rm HK}$), each corresponding to a different parameter set (Table~\ref{tab_param}). This leads to 11421 time series, or 22842 when the two levels for Tspot are considered for each inclination and observable, hence a total of 228420 realizations for each observable. 
Inclinations take values between 0$^{\circ}$ (pole-on) and 90$^{\circ}$ (edge-on), with a step of 10$^{\circ}$. 
Fig~\ref{recap_param} shows a summary of all parameters.

\subsection{Temporal sampling and duration}

\subsubsection{Reference sampling}

Time series must have a sufficiently long duration to allow us to test analysis methods, and they must cover at least a cycle period. To keep it reasonable, however, we imposed a maximum of 15 years (which is just above the maximum cycle period we considered). We then simulated an integer number of cycles, choosing the maximum number that would lead to a duration shorter than 15 years. 

The time step was one day on average, but as in \cite{borgniet15}, we added a small random departure (within $\pm$ 4 hours) from the regular sampling to mimic a realistic sampling.

\subsection{Addition of short-term variablity in RV}

\subsubsection{Principle}

To produce realistic RV time series that include all contributions at various timescales, we added the contribution of oscillations, granulation, and supergranulation, as was done in \cite{dumusque16}. We call these three contributions to RV the OGS signal hereafter.
The principle is the following. For each spectral type and each variable, we computed one time series that covered 15 years and had a time step of 30 seconds. From this, the time series corresponding to a given sampling can be extracted.  To produce a time series like this, we computed the inverse Fourier transform of the power spectrum as a function of the frequency $\nu$, P($\nu$), for each of these contributions.
The parameters describing the power depend on the spectral type. We also computed smoothed time series (with a bin of one hour) to simulate the effect of long exposure times. In the following, we mostly use such long-time exposure time series, assuming some good observing conditions. The series with no smoothing may be used for a comparison with observations that were made with short exposure times, however. In practice, a long-duration (15 years) time series was produced for each spectral type, with a time step of 30 seconds; for a given time series in our grid, we extracted either the instantaneous values or the one-hour average corresponding to the same sampling.

We can also add a white noise of 0.6~m/s to simulate instrumental noise (this value corresponds to typical uncertainties on individual measurement from HARPS data for G stars).  We considered four types of time series in our analysis: 
\begin{itemize}
\item{1: original RV time series caused by magnetic activity. These are useful to study only the activity contribution.}
\item{2: original RV time series plus oscillation, granulation, or supergranulation signal (no smoothing). This is useful to compare with observations under ideal conditions (assuming the instrumental noise was totally corrected for).   }
\item{3: original RV time series plus oscillation,  granulation, or supergranulation signal (no smoothing) plus instrumental noise}.
\item{4: original RV time series plus oscillation, granulation, or supergranulation signal (one-hour smoothing) plus instrumental noise}.
\end{itemize}

\begin{figure}
\includegraphics{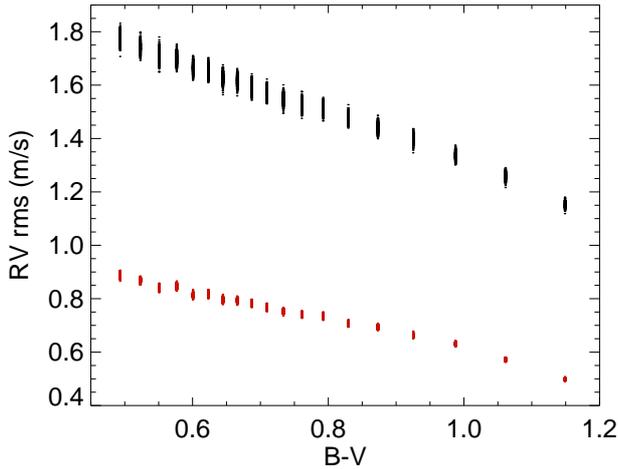}
\caption{
RV jitter due to the oscillation, granulation, and supergranulation vs. B-V for instantaneous values (black) and averaged over one hour (red). 
}
\label{ogs}
\end{figure}

\subsubsection{Oscillations}

We used the following power function for the oscillations: 
\begin{equation}
P(\nu)=A \times  {\rm e}^{-(\nu-\nu_0)^2/2/\Gamma^2}
,\end{equation}

which describes the mode envelopes and not the individual modes themselves \cite[e.g.,][]{kallinger14}. We adopted the following scaling laws for different types of stars:
\begin{equation}
A=(T_{\rm eff}/T_{\rm eff,\odot})^4 \times (R/R_{\odot})^2 / (M/M_{\odot})^{0.7}
\end{equation}
from \cite{kjeldsen95}, with the adaptation of \cite{samadi07} for the exponent, where $A$ is relative to the solar value, 


\begin{equation}
\nu_0=(M/M_{\odot}) / (R/R_{\odot})^2  / \sqrt{T_{\rm eff}/T_{\rm eff,\odot}}
,\end{equation}
from \cite{bedding03}, where $\nu_0$ is relative to the solar value, and 

\begin{equation}
\Gamma=\sqrt{ (M/M_{\odot)} } / (R/R_{\odot})^{1.5}
\end{equation}
from \cite{kippenhahn90} and \cite{belkacem13}, where $\Gamma$ is relative to the solar value. 

These laws were scaled with the following solar values: $A_{\odot}$=200 (m/s)$^2$/Hz \cite[which provides an amplitude of the power that agrees well with the observed power, e.g., from][]{davies14}, $\nu_{0,\odot}$=3140.10$^{-6}$ Hz, and $\Gamma_{\odot}$=361.10$^{-6}$ Hz \cite[both from][]{kallinger14}.

\subsubsection{Granulation}

We used the following power function for the granulation signal: 
\begin{equation}
P(\nu)=A/(1+(\tau \nu)^\beta)
\end{equation}

from \cite{harvey84}.
The power spectrum in RV of a few stars was analyzed \cite[][]{dumusque11b}, but the sample is not large enough to derive a proper trend.  We therefore used a scaling derived from the numerical simulation of \cite{beeck13a} to obtain the scaling of $A$ and $\tau$ by fitting a linear law on their results: 
\begin{equation}
A = (0.3+6.323 10^{-4}\times (T_{\rm eff}-4594))
,\end{equation}
and
\begin{equation}
\tau=-2.831+1.574 10^{-3} \times T_{\rm eff}
\end{equation}
 are normalized by the same amplitude and $\tau$ respectively, computed for 5784 K (G2).
$\beta$ was kept to the solar value. The solar values were derived from a fit on the simulated time series produced in \cite{meunier15}:
$A_{\odot}$=154 (m/s)$^2$/Hz, $\tau_{\odot}$=2781 sec, and $\beta_{\odot}$=1.97.

\subsubsection{Supergranulation}

The formula for the supergranulation power is similar to equation 10 for granulation. Supergranulation seems to be present in stars other than the Sun \cite[][]{dumusque11b}, but the statistics is not sufficient to describe the parameters as a function of spectral type. Given the lack of knowledge on stellar supergranulation and  because it is likely to be related to the granulation pattern and amplitude \cite[e.g.,][]{roudier16}, we used the granulation scaling relation. We considered the solar time series simulated in \cite{meunier15}, which correspond to intermediate parameters between the two extremes the authors evaluated (supergranulation is less strongly constrained than granulation), and then fitted them, which gives the following parameters: A=43000 (m/s)$^2$/Hz and $\tau$=1.1 10$^{_6}$ sec. $\beta$ was kept fixed to the granulation value. 

\subsubsection{RV jitter caused by the OGS signal}

The rms RV produced by the OGS signal alone is shown in Fig.~\ref{ogs} as a function of B-V.  
It ranges from 1.8 to 1.1 m/s for stars between F6 and K4. After averaging over one hour, the values are lower than 1 m/s. They lie between 0.9 and 0.5 m/s from F6 to K4. 

\section{Chromospheric emission - calibrating the spot number}

\subsection{Objectives and principle}

The laws described in Sects.~2 and 3 depend on LogR'$_{\rm HK}$, which is an observable. However, the input parameters of our simulations are the number of spots (Sect.~2.1.1). We therefore need to know how many spots to inject if we wish to reach a given activity level that is described by its average LogR'$_{\rm HK}$. For this purpose, we need two elements: a chromospheric emission model that uses the input parameters of our simulations, and a calibration law relating LogR'$_{\rm HK}$ and spot number. This model will also be very useful after the simulations are performed to determine to which LogR'$_{\rm HK}$ they correspond as well: the exact average LogR'$_{\rm HK}$ of simulation may be slightly different from the one in Fig.~\ref{bv_logrphk}, but our objective is that it should be close (so that the input parameters we have used for that particular simulation are valid).

Assuming a solar chromospheric emission model, we therefore performed a series of simulations with constant activity levels, which will give the typical contribution per injected spot to the S-index. This was made for a constant number of spots\footnote{ We note that in this paper the spot number is a true spot number, that is the number of individuals spots we actually inject. This is different from \cite{borgniet15}, in which we considered the solar Wolf number, which is a combination of the number of spots and the number of spots groups, which needed a conversion into a number of individual spots to be injected. This complexity is not necessary here as this was a purely solar approach which does not make sense for other stars. } 
on eight-year time series. The average was then made on inclination (because the same structures were used for all inclinations, they correspond to the same inputs in terms of spot number), $\theta_{\rm max}$ (very small variation), and spectral type (no trend observed, which is expected because we used the same law to compute the plage contribution). 
We note that this calibration depends on the plage-to-spot size ratio: we kept it constant in this paper, but if it were to vary, new calibrations are required. The same is true if the size distribution of spots changes.

\subsection{Principle of the chromospheric emission model}

A necessary step is therefore to implement a model providing an S-index (and then LogR'$_{\rm HK}$) based on a list of plages and network features at each time step. The full model is described in Appendix D. We provide here the general principles. 
The model is based on the work of \cite{meunier18a} for the Sun, and includes three contributions: 1/ the basal flux (when no activity is present, determined in Sect.~2.3.1); 2/ the contribution of plages and network structures, with a law that depends on their size, following \cite{harvey99}; 3/ the contribution of the quiet star ("quiet" here means outside active regions and network, hereafter QS). This last contribution is important because it must vary from one star to the other: if it were kept constant, it would be impossible to observe stars with variability while having an average activity level below that of the quiet Sun because the level at solar minimum (corresponding to no structures) is significantly above the basal level. This is discussed in more detail in Appendix D. 
We propose that this contribution depends on the average activity level of the star: the more active the star, the stronger the (weak) magnetic field in the quiet star, and the larger the contribution of the quiet star to the chromospheric emission. The exact choice of the QS contribution will affect the number of spots, and most especially, the number at cycle minimum.



For a given LogR'$_{\rm HK}$, we computed the S-index, from which the basal flux was removed, as well as the QS contribution for that LogR'$_{\rm HK}$ (see Appendix D.1). The resulting flux was divided by the typical flux per structure, providing the number of spots to inject to obtain the LogR'$_{\rm HK}$ we wish to reach. 
The resulting calibration works very well, as we show in the next section. 


\subsection{List of the different time series}

\begin{table}
\caption{Time series}
\label{tab_list}
\begin{center}
\renewcommand{\footnoterule}{}  
\begin{tabular}{ll}
\hline
 & Variable \\ \hline 
RV & \\
   & RVspot$_1$, RVspot$_2$ \\
   & RVplage  \\
   & RVconv  \\
   & OSG  \\
   & Inst. Noise  \\
\hline
Photometry &  \\
   & Ispot$_1$, Ispot$_2$  \\
   & Iplage  \\
\hline
Astrometry &  \\
   & Xspot$_1$, Xspot$_2$  \\
   & Yspot$_1$, Yspot$_2$  \\
   & Xplage  \\
   & Yplage  \\
\hline
Chromospheric emission &  \\
   & S-index  \\
   & LogR'$_{\rm HK}$  \\
\hline
Other variables &    \\
   & ff spot  \\
   & ff plage  \\
   & nb spot  \\
   & nb plage  \\
   & nb network  \\
\hline
\end{tabular}
\end{center}
\tablefoot{ff and nb are the apparent filling factor and number of structures, respectively; they are not observables. The other variables are either observables or can produce observables when several of these variables are combined.  Subscripts 1 and 2 refer to the two laws for $\Delta$spot. 
} 
\end{table}

Table~\ref{tab_list} shows the list of time series that we produced during the simulations. We recall that except for the number of structures, plage refers to all bright structures, from large structures in active regions to the  smallest structures in the network.


\section{LogR'$_{\rm HK}$ behavior}


\subsection{Comparison between objective and realization}

\begin{figure}
\includegraphics{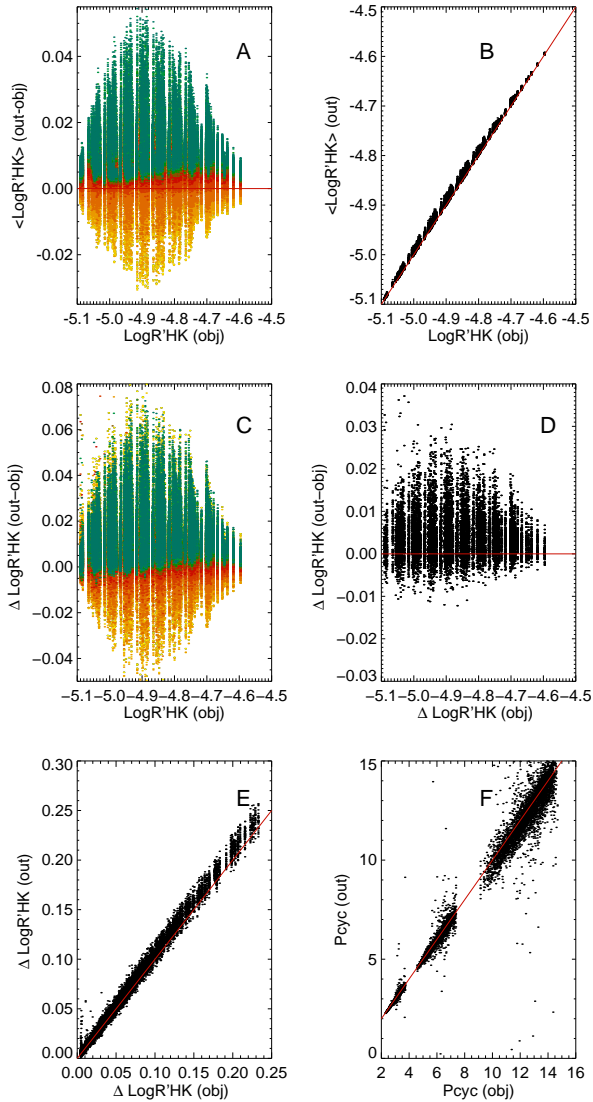}
\caption{
{\it Panel A:} LogR'$_{\rm HK,out}$-LogR'$_{\rm HK,obj}$ vs. LogR'$_{\rm HK,obj}$ for all simulations. The color indicates stellar inclination from 0$^{\circ}$ (pole-on, yellow) to 90$^{\circ}$ (edge-on, blue). 
{\it Panel B:} LogR'$_{\rm HK,out}$ averaged over all inclinations vs. LogR'$_{\rm HK,obj}$. The solid red line indicates the y=x linear function. 
{\it Panel C:} $\Delta$LogR'$_{\rm HK,out}$-$\Delta$LogR'$_{\rm HK,obj}$ vs. LogR'$_{\rm HK,obj}$, where $\Delta$ represents the amplitude of the cycle, for different inclinations (the color code is similar to panel A).
{\it Panel D:} $\Delta$LogR'$_{\rm HK,out}$-$\Delta$LogR'$_{\rm HK,obj}$ vs. LogR'$_{\rm HK,obj}$, where $\Delta$ represents the amplitude of the cycle, after averaging the simulations made for inclination of 40$^{\circ}$ and 50$^{\circ}$.
{\it Panel E:} $\Delta$LogR'$_{\rm HK,out}$ vs. $\Delta$LogR'$_{\rm HK,obj}$, after averaging for inclination of 40$^{\circ}$ and 50$^{\circ}$.
The solid red line indicates the y=x linear function. 
{\it Panel F:} Cycle period (in years) derived from a fit on smoothed LogR'$_{\rm HK}$ time series vs. the prescribed cycle period. 
}
\label{logrphk_check}
\end{figure}


As explained in Sects.~2 and 4, we wish to simulate time series with a given average activity level, as well as a certain cycle amplitude (in LogR'$_{\rm HK}$), that is, LogR'$_{\rm HK,obj}$ and $\Delta$LogR'$_{\rm HK,obj}$. A calibration was necessary to achieve this goal (Sect.~4.2), therefore we must check that the simulations behave as planned. We computed the average LogR'$_{\rm HK,out}$ for each time series, where "out" stands for the output LogR'HK from the simulations. The properties of this realized LogR'HK ("out") time series was compared to the targeted LogR'HK ("obj") from the grid of parameters from Sect.~2.3. The amplitude, $\Delta$LogR'$_{\rm HK,out}$, was computed using a sinusoidal fit of each smoothed LogR'$_{\rm HK}$ time series.

The two upper panels (A and B) in Fig.~\ref{logrphk_check} compare the LogR'$_{\rm HK,obj}$ and LogR'HK$_{\rm HK,out}$. We observe a strong inclination effect on LogR'$_{\rm HK}$; departures from the average are within 0.05 dex. However, on average, the difference between the expected value and the final value is much smaller when this inclination dependence is removed (below 0.01 typically). The differences are smaller than the typical uncertainties on LogR'$_{\rm HK}$ values as estimated by \cite{radick18}, of about 0.06 dex, and our LogR'$_{\rm HK}$ should not be considered to be more precise than this in absolute value (although for a given simulation, the relative variability will be much more precise, of course). 

In Appendix D, we mention the possibility of a trend in chromospheric emission versus T$_{\rm eff}$. We did not include this trend because it is still very uncertain.
When it is applied, the difference is small; the largest difference for our lowest mass stars is about 0.03 dex. 
If it is real, we would need fewer spots for the stars with the lowest mass than are included in the present simulations to reach a given objective because the emission for a given plage would be higher. 

A fit of the time series with a sinusoidal provides an estimate of the amplitude of the cycle and of its period, which can be compared to expectations. 
Panel C in Fig.~\ref{logrphk_check} shows the inclination effect on the cycle amplitude, while 
panels D and E in Fig.~\ref{logrphk_check} compare $\Delta$LogR'$_{\rm HK,obj}$ and  $\Delta$LogR'$_{\rm HK,out}$ for average inclinations. The cycle amplitudes from the simulations are also very close to the expected ones. 
Finally, the last plot (panel F) compares cycle periods, which for most simulations are in good agreement. There are a few outliers, but these are mostly due to low-amplitude simulations, for which the measurement itself is not reliable.  

We conclude that after averaging, the average LogR'$_{\rm HK}$, the amplitude, and period of the cycle agree with the input parameters. The  inclination effect is discussed in the next section. 

\subsection{Dependence on inclination}

We observe a strong inclination effect on the average LogR'$_{\rm HK}$, with a stronger value for an edge-on than for a pole-on configuration. This is in agreement with the results of \cite{shapiro14}, which were based on simulations with a simpler model of the chromospheric emission (no structures or size dependence). \cite{knaack01} obtained a much weaker dependence, probably because they used a model that did not take all parameters into account. 
The inclination effect is also strong on the long-term amplitude in LogR'$_{\rm HK}$, although it presents a large dispersion. For the solar $\theta_{\rm max}$, the amplitude is larger for the edge-on configuration, with a difference of about 20-40\% depending on the simulation. This has been observed by \cite{knaack01}. For larger $\theta_{\rm max}$, the difference is smaller, with aa slight predominance of larger amplitude when edge-on for $\theta_{\rm max,\odot}$+10$^{\circ}$ and a reversal for $\theta_{\rm max,\odot}$+20$^{\circ}$ (with a large dispersion and difference occasionally up to 20\%).  


\subsection{Example of RV and LogR'$_{\rm HK}$ time series}

\begin{figure}
\includegraphics{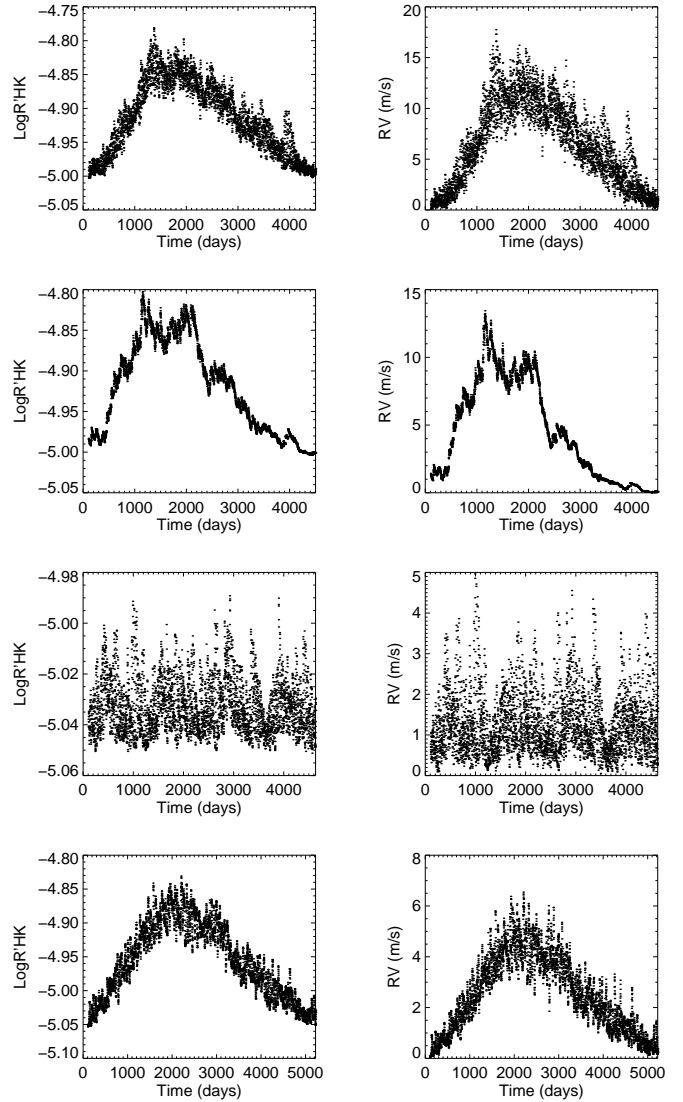}
\caption{
{\it First line:}  LogR'$_{\rm HK}$ (left) and RV (right) vs. time for a moderately active G2 star and an inclination of 90$^{\circ}$. 
{\it Second line:} Same for an inclination of 0$^{\circ}$. 
{\it Third line:} Same for a quiet G2 star and an inclination of 90$^{\circ}$. 
{\it Fourth line:} Same for a moderately active K2 star and an inclination of 90$^{\circ}$. 
}
\label{exemple_rv_ca}
\end{figure}

Fig.~\ref{exemple_rv_ca} shows a few examples of time series for a small sample of spectral types, activity levels, and inclinations in chromospheric emission and radial velocity. The different activity levels correspond to different ages. When the law derived by \cite{mamajek08} to relate rotation and age is assumed, the first two panels would correspond to an age of 3.3 Gyr, the third panel to 4.3 Gyr, and the fourth to 8.2 Gyr. This shows a good similarity between RV and  LogR'$_{\rm HK}$ long-term variations. It is also possible to obtain relatively flat curves for the lowest cycle amplitudes.

\section{Conclusion}

We have proposed a  model to produce realistic time series of different variables (RV, photometry, astrometry, and chromospheric emission) that represent complex activity patterns for a wide range of stars.  We have described the model in detail:  
A specificity of our simulations is that we use consistent parameter sets for a wide range of stars, that is, old F6-K4 star with different activity levels.

Our very large set of time series will be analyzed  in detail in subsequent papers. We will compare the RV jitter between simulations and observations, analyze the effect of parameters on the RV jitter, and use this RV jitter to predict the detectability of exoplanets as a function of B-V and LogR$_{\rm HK}$ \cite[][]{meunier18c}. The detailed relationship between RV and LogR'$_{\rm HK}$ will be studied in order to understand why the corrections of RV time series using a linear function of LogR'$_{\rm HK}$ are limited \cite[][]{meunier18d}. RV times series will be further analyzed to produce detection limits by taking the frequency behavior of the stellar variability into account. Finally, a similar analysis will be made for astrometry. The effect of oscillation, granulation, and supergranulation will also be studied in more detail as a function of spectral type and activity level. 

These time series are a good tool to provide clues to help interpret stellar variability from brightness time series: this is crucial because there are many degeneracies and biases, and synthetic time series are useful to determine the effect of the different parameters. They can also be used to test new methods, not only a correcting method for purposes of exoplanet detection, but of stellar activity analysis.  


When these globally consistent parameter sets are built, the main limitation in our opinions is the poorly constrained QS contribution to the chromospheric emission. We have made two strong assumptions because our knowledge is incomplete. First, we have neglected the variation of this contribution with time, although we expect a small variation with a complex pattern from our solar study \cite[][]{meunier18a} (competition between stronger magnetic field at cycle maximum, but also a lower surface coverage). Second, we imposed that the number of spots at cycle minimum varies within a small range (and is small). We cannot exclude that for very active stars, for example, there could be a trend of a lower QS contribution and larger spot number at cycle minimum. This is not constrained,  however, although we know that the Sun, which lies in the middle of our grid, has very few spots at cycle minimum and therefore conforms to our assumption. The only way  to go beyond this limitation would be to better understand this QS contribution over the cycle and for different activity levels, most likely from dedicated simulations (MHD or using flux tubes down to very small spatial scales).  

\begin{acknowledgements}

This work has been funded by the Universit\'e Grenoble Alpes project call "Alpes Grenoble Innovation Recherche (AGIR)" and the ANR GIPSE ANR-14-CE33-0018.
We are very grateful to Charlotte Norris, who has provided us the plage contrasts used in this work prior the publication of her thesis. We thank Fr\'ed\'eric Baudin and Sasha Brun for useful discussions about some aspects of this work. We thank Pascal Rubini for his work of convert the simulation code into C++. 
We thank the anonymous referee for his/her useful comments which helped to improve the paper. 

\end{acknowledgements}

\bibliographystyle{aa}
\bibliography{biblio}

\begin{thebibliography}{115}
\expandafter\ifx\csname natexlab\endcsname\relax\def\natexlab#1{#1}\fi

\bibitem[{{Arriagada}(2011)}]{arriagada11}
{Arriagada}, P. 2011, \apj, 734, 70

\bibitem[{{Baliunas} {et~al.}(1998){Baliunas}, {Donahue}, {Soon}, \&
  {Henry}}]{baliunas98}
{Baliunas}, S.~L., {Donahue}, R.~A., {Soon}, W., \& {Henry}, G.~W. 1998, in
  Astronomical Society of the Pacific Conference Series, Vol. 154, Cool Stars,
  Stellar Systems, and the Sun, ed. R.~A. {Donahue} \& J.~A. {Bookbinder}, 153

\bibitem[{{Baliunas} {et~al.}(1996){Baliunas}, {Nesme-Ribes}, {Sokoloff}, \&
  {Soon}}]{baliunas96}
{Baliunas}, S.~L., {Nesme-Ribes}, E., {Sokoloff}, D., \& {Soon}, W.~H. 1996,
  \apj, 460, 848

\bibitem[{{Balona} \& {Abedigamba}(2016)}]{balona16}
{Balona}, L.~A. \& {Abedigamba}, O.~P. 2016, \mnras, 461, 497

\bibitem[{{Barnes} {et~al.}(2005){Barnes}, {Collier Cameron}, {Donati},
  {James}, {Marsden}, \& {Petit}}]{barnes05}
{Barnes}, J.~R., {Collier Cameron}, A., {Donati}, J.-F., {et~al.} 2005, \mnras,
  357, L1

\bibitem[{{Baumann} \& {Solanki}(2005)}]{baumann05}
{Baumann}, I. \& {Solanki}, S.~K. 2005, \aap, 443, 1061

\bibitem[{{Bedding} \& {Kjeldsen}(2003)}]{bedding03}
{Bedding}, T.~R. \& {Kjeldsen}, H. 2003, \pasa, 20, 203

\bibitem[{{Beeck} {et~al.}(2013){Beeck}, {Cameron}, {Reiners}, \&
  {Sch{\"u}ssler}}]{beeck13a}
{Beeck}, B., {Cameron}, R.~H., {Reiners}, A., \& {Sch{\"u}ssler}, M. 2013,
  \aap, 558, A48

\bibitem[{{Beeck} {et~al.}(2015){Beeck}, {Sch{\"u}ssler}, {Cameron}, \&
  {Reiners}}]{beeck15}
{Beeck}, B., {Sch{\"u}ssler}, M., {Cameron}, R.~H., \& {Reiners}, A. 2015,
  \aap, 581, A42

\bibitem[{{Belkacem} {et~al.}(2013){Belkacem}, {Samadi}, {Mosser}, {Goupil}, \&
  {Ludwig}}]{belkacem13}
{Belkacem}, K., {Samadi}, R., {Mosser}, B., {Goupil}, M.-J., \& {Ludwig}, H.-G.
  2013, in Astronomical Society of the Pacific Conference Series, Vol. 479,
  Progress in Physics of the Sun and Stars: A New Era in Helio- and
  Asteroseismology, ed. H.~{Shibahashi} \& A.~E. {Lynas-Gray}, 61

\bibitem[{{Berdyugina}(2005)}]{berd05}
{Berdyugina}, S.~V. 2005, Living Reviews in Solar Physics, 2, 8

\bibitem[{{B{\"o}hm-Vitense}(2007)}]{bohm07}
{B{\"o}hm-Vitense}, E. 2007, \apj, 657, 486

\bibitem[{{Borgniet} {et~al.}(2015){Borgniet}, {Meunier}, \&
  {Lagrange}}]{borgniet15}
{Borgniet}, S., {Meunier}, N., \& {Lagrange}, A.-M. 2015, \aap, 581, A133

\bibitem[{{Boyajian} {et~al.}(2013){Boyajian}, {von Braun}, {van Belle},
  {Farrington}, {Schaefer}, {Jones}, {White}, {McAlister}, {ten Brummelaar},
  {Ridgway}, {Gies}, {Sturmann}, {Sturmann}, {Turner}, {Goldfinger}, \&
  {Vargas}}]{boyajian13}
{Boyajian}, T.~S., {von Braun}, K., {van Belle}, G., {et~al.} 2013, \apj, 771,
  40

\bibitem[{{Boyajian} {et~al.}(2012){Boyajian}, {von Braun}, {van Belle},
  {McAlister}, {ten Brummelaar}, {Kane}, {Muirhead}, {Jones}, {White},
  {Schaefer}, {Ciardi}, {Henry}, {L{\'o}pez-Morales}, {Ridgway}, {Gies}, {Jao},
  {Rojas-Ayala}, {Parks}, {Sturmann}, {Sturmann}, {Turner}, {Farrington},
  {Goldfinger}, \& {Berger}}]{boyajian12}
{Boyajian}, T.~S., {von Braun}, K., {van Belle}, G., {et~al.} 2012, \apj, 757,
  112

\bibitem[{{Brandt} \& {Solanki}(1990)}]{BS90}
{Brandt}, P.~N. \& {Solanki}, S.~K. 1990, \aap, 231, 221

\bibitem[{{Brun} {et~al.}(2017){Brun}, {Strugarek}, {Varela}, {Matt},
  {Augustson}, {Emeriau}, {DoCao}, {Brown}, \& {Toomre}}]{brun17}
{Brun}, A.~S., {Strugarek}, A., {Varela}, J., {et~al.} 2017, \apj, 836, 192

\bibitem[{{Cegla} {et~al.}(2018){Cegla}, {Watson}, {Shelyag}, {Chaplin},
  {Davies}, {Mathioudakis}, {Palumbo}, {Saar}, \& {Haywood}}]{cegla18}
{Cegla}, H.~M., {Watson}, C.~A., {Shelyag}, S., {et~al.} 2018, \apj, 866, 55

\bibitem[{{Claret} \& {Hauschildt}(2003)}]{claret03}
{Claret}, A. \& {Hauschildt}, P.~H. 2003, \aap, 412, 241

\bibitem[{{Collier Cameron}(2007)}]{colliercameron07}
{Collier Cameron}, A. 2007, Astronomische Nachrichten, 328, 1030

\bibitem[{{Cuntz} {et~al.}(1999){Cuntz}, {Rammacher}, {Ulmschneider},
  {Musielak}, \& {Saar}}]{cuntz99}
{Cuntz}, M., {Rammacher}, W., {Ulmschneider}, P., {Musielak}, Z.~E., \& {Saar},
  S.~H. 1999, \apj, 522, 1053

\bibitem[{{Das Chagas} {et~al.}(2016){Das Chagas}, {Bravo}, {Costa}, {Ferreira
  Lopes}, {Silva Sobrinho}, {Paz-Chinch{\'o}n}, {Le{\~a}o}, {Valio}, {de
  Freitas}, {Canto Martins}, {Lanza}, \& {De Medeiros}}]{daschagas16}
{Das Chagas}, M.~L., {Bravo}, J.~P., {Costa}, A.~D., {et~al.} 2016, \mnras,
  463, 1624

\bibitem[{{Davies} {et~al.}(2014){Davies}, {Chaplin}, {Elsworth}, \&
  {Hale}}]{davies14}
{Davies}, G.~R., {Chaplin}, W.~J., {Elsworth}, Y., \& {Hale}, S.~J. 2014,
  \mnras, 441, 3009

\bibitem[{{Desort} {et~al.}(2007){Desort}, {Lagrange}, {Galland}, {Udry}, \&
  {Mayor}}]{desort07}
{Desort}, M., {Lagrange}, A.-M., {Galland}, F., {Udry}, S., \& {Mayor}, M.
  2007, \aap, 473, 983

\bibitem[{{Donahue} {et~al.}(1996){Donahue}, {Saar}, \& {Baliunas}}]{donahue96}
{Donahue}, R.~A., {Saar}, S.~H., \& {Baliunas}, S.~L. 1996, \apj, 466, 384

\bibitem[{{Dumusque}(2016)}]{dumusque16}
{Dumusque}, X. 2016, \aap, 593, A5

\bibitem[{{Dumusque} {et~al.}(2017){Dumusque}, {Borsa}, {Damasso},
  {D{\'{\i}}az}, {Gregory}, {Hara}, {Hatzes}, {Rajpaul}, {Tuomi}, {Aigrain},
  {Anglada-Escud{\'e}}, {Bonomo}, {Bou{\'e}}, {Dauvergne}, {Frustagli},
  {Giacobbe}, {Haywood}, {Jones}, {Laskar}, {Pinamonti}, {Poretti}, {Rainer},
  {S{\'e}gransan}, {Sozzetti}, \& {Udry}}]{dumusque17}
{Dumusque}, X., {Borsa}, F., {Damasso}, M., {et~al.} 2017, \aap, 598, A133

\bibitem[{{Dumusque} {et~al.}(2015){Dumusque}, {Glenday}, {Phillips},
  {Buchschacher}, {Collier Cameron}, {Cecconi}, {Charbonneau}, {Cosentino},
  {Ghedina}, {Latham}, {Li}, {Lodi}, {Lovis}, {Molinari}, {Pepe}, {Udry},
  {Sasselov}, {Szentgyorgyi}, \& {Walsworth}}]{dumusque15}
{Dumusque}, X., {Glenday}, A., {Phillips}, D.~F., {et~al.} 2015, \apjl, 814,
  L21

\bibitem[{{Dumusque} {et~al.}(2011){Dumusque}, {Udry}, {Lovis}, {Santos}, \&
  {Monteiro}}]{dumusque11b}
{Dumusque}, X., {Udry}, S., {Lovis}, C., {Santos}, N.~C., \& {Monteiro},
  M.~J.~P.~F.~G. 2011, \aap, 525, A140

\bibitem[{{Fawzy} {et~al.}(2002{\natexlab{a}}){Fawzy}, {Rammacher},
  {Ulmschneider}, {Musielak}, \& {St{\c e}pie{\'n}}}]{fawzy02a}
{Fawzy}, D., {Rammacher}, W., {Ulmschneider}, P., {Musielak}, Z.~E., \& {St{\c
  e}pie{\'n}}, K. 2002{\natexlab{a}}, \aap, 386, 971

\bibitem[{{Fawzy} {et~al.}(2002{\natexlab{b}}){Fawzy}, {St{\c e}pie{\'n}},
  {Ulmschneider}, {Rammacher}, \& {Musielak}}]{fawzy02c}
{Fawzy}, D., {St{\c e}pie{\'n}}, K., {Ulmschneider}, P., {Rammacher}, W., \&
  {Musielak}, Z.~E. 2002{\natexlab{b}}, \aap, 386, 994

\bibitem[{{Fawzy} {et~al.}(2002{\natexlab{c}}){Fawzy}, {Ulmschneider}, {St{\c
  e}pie{\'n}}, {Musielak}, \& {Rammacher}}]{fawzy02b}
{Fawzy}, D., {Ulmschneider}, P., {St{\c e}pie{\'n}}, K., {Musielak}, Z.~E., \&
  {Rammacher}, W. 2002{\natexlab{c}}, \aap, 386, 983

\bibitem[{{Garc{\'{\i}}a} {et~al.}(2014){Garc{\'{\i}}a}, {Ceillier},
  {Salabert}, {Mathur}, {van Saders}, {Pinsonneault}, {Ballot}, {Beck},
  {Bloemen}, {Campante}, {Davies}, {do Nascimento}, {Mathis}, {Metcalfe},
  {Nielsen}, {Su{\'a}rez}, {Chaplin}, {Jim{\'e}nez}, \& {Karoff}}]{garcia14}
{Garc{\'{\i}}a}, R.~A., {Ceillier}, T., {Salabert}, D., {et~al.} 2014, \aap,
  572, A34

\bibitem[{{Gray}(2005)}]{gray05}
{Gray}, D.~F. 2005, {The Observation and Analysis of Stellar Photospheres}

\bibitem[{{Gray} {et~al.}(2006){Gray}, {Corbally}, {Garrison}, {McFadden},
  {Bubar}, {McGahee}, {O'Donoghue}, \& {Knox}}]{gray06}
{Gray}, R.~O., {Corbally}, C.~J., {Garrison}, R.~F., {et~al.} 2006, \aj, 132,
  161

\bibitem[{{Gray} {et~al.}(2003){Gray}, {Corbally}, {Garrison}, {McFadden}, \&
  {Robinson}}]{gray03}
{Gray}, R.~O., {Corbally}, C.~J., {Garrison}, R.~F., {McFadden}, M.~T., \&
  {Robinson}, P.~E. 2003, \aj, 126, 2048

\bibitem[{{Hall}(1991)}]{hall91}
{Hall}, D.~S. 1991, in Lecture Notes in Physics, Berlin Springer Verlag, Vol.
  380, IAU Colloq. 130: The Sun and Cool Stars. Activity, Magnetism, Dynamos,
  ed. I.~{Tuominen}, D.~{Moss}, \& G.~{R{\"u}diger}, 353

\bibitem[{{Hall} {et~al.}(2009){Hall}, {Henry}, {Lockwood}, {Skiff}, \&
  {Saar}}]{hall09}
{Hall}, J.~C., {Henry}, G.~W., {Lockwood}, G.~W., {Skiff}, B.~A., \& {Saar},
  S.~H. 2009, \aj, 138, 312

\bibitem[{{Hall} {et~al.}(2019)}]{hall19}
{Hall}, J.~C. {et~al.} 2019, in preparation

\bibitem[{{Harvey}(1984)}]{harvey84}
{Harvey}, J.~W. 1984, in Probing the depths of a Star: the study of Solar
  oscillation from space, ed. R. W. Noyes, \& E. J. Rhodes Jr., JPL, 400, 327

\bibitem[{{Harvey} \& {White}(1999)}]{harvey99}
{Harvey}, K.~L. \& {White}, O.~R. 1999, \apj, 515, 812

\bibitem[{{Hatzes}(2002)}]{hatzes02}
{Hatzes}, A.~P. 2002, Astronomische Nachrichten, 323, 392

\bibitem[{{Haywood} {et~al.}(2016){Haywood}, {Collier Cameron}, {Unruh},
  {Lovis}, {Lanza}, {Llama}, {Deleuil}, {Fares}, {Gillon}, {Moutou}, {Pepe},
  {Pollacco}, {Queloz}, \& {S{\'e}gransan}}]{haywood16}
{Haywood}, R.~D., {Collier Cameron}, A., {Unruh}, Y.~C., {et~al.} 2016, \mnras,
  457, 3637

\bibitem[{{Henry} {et~al.}(1996){Henry}, {Soderblom}, {Donahue}, \&
  {Baliunas}}]{henry96}
{Henry}, T.~J., {Soderblom}, D.~R., {Donahue}, R.~A., \& {Baliunas}, S.~L.
  1996, \aj, 111

\bibitem[{{Herrero} {et~al.}(2016){Herrero}, {Ribas}, {Jordi}, {Morales},
  {Perger}, \& {Rosich}}]{herrero16}
{Herrero}, E., {Ribas}, I., {Jordi}, C., {et~al.} 2016, \aap, 586, A131

\bibitem[{{I{\c s}{\i}k} {et~al.}(2011){I{\c s}{\i}k}, {Schmitt}, \&
  {Sch{\"u}ssler}}]{isik11}
{I{\c s}{\i}k}, E., {Schmitt}, D., \& {Sch{\"u}ssler}, M. 2011, \aap, 528, A135

\bibitem[{{Isaacson} \& {Fischer}(2010)}]{isaacson10}
{Isaacson}, H. \& {Fischer}, D. 2010, \apj, 725, 875

\bibitem[{{Jenkins} {et~al.}(2008){Jenkins}, {Jones}, {Pavlenko}, {Pinfield},
  {Barnes}, \& {Lyubchik}}]{jenkins08}
{Jenkins}, J.~S., {Jones}, H.~R.~A., {Pavlenko}, Y., {et~al.} 2008, \aap, 485,
  571

\bibitem[{{Jenkins} {et~al.}(2011){Jenkins}, {Murgas}, {Rojo}, {Jones},
  {Day-Jones}, {Jones}, {Clarke}, {Ruiz}, \& {Pinfield}}]{jenkins11}
{Jenkins}, J.~S., {Murgas}, F., {Rojo}, P., {et~al.} 2011, \aap, 531, A8

\bibitem[{{Kallinger} {et~al.}(2014){Kallinger}, {De Ridder}, {Hekker},
  {Mathur}, {Mosser}, {Gruberbauer}, {Garc{\'{\i}}a}, {Karoff}, \&
  {Ballot}}]{kallinger14}
{Kallinger}, T., {De Ridder}, J., {Hekker}, S., {et~al.} 2014, \aap, 570, A41

\bibitem[{{Kippenhahn} \& {Weigert}(1990)}]{kippenhahn90}
{Kippenhahn}, R. \& {Weigert}, A. 1990, {Stellar Structure and Evolution}, 192

\bibitem[{{Kjeldsen} \& {Bedding}(1995)}]{kjeldsen95}
{Kjeldsen}, H. \& {Bedding}, T.~R. 1995, \aap, 293, 87

\bibitem[{{Knaack} {et~al.}(2001){Knaack}, {Fligge}, {Solanki}, \&
  {Unruh}}]{knaack01}
{Knaack}, R., {Fligge}, M., {Solanki}, S.~K., \& {Unruh}, Y.~C. 2001, \aap,
  376, 1080

\bibitem[{{Komm} {et~al.}(1993){Komm}, {Howard}, \& {Harvey}}]{komm93}
{Komm}, R.~W., {Howard}, R.~F., \& {Harvey}, J.~W. 1993, \solphys, 147, 207

\bibitem[{{K{\"u}ker} \& {R{\"u}diger}(2011)}]{kuker11}
{K{\"u}ker}, M. \& {R{\"u}diger}, G. 2011, Astronomische Nachrichten, 332, 933

\bibitem[{{K{\"u}ker} {et~al.}(2011){K{\"u}ker}, {R{\"u}diger}, \&
  {Kitchatinov}}]{kuker11b}
{K{\"u}ker}, M., {R{\"u}diger}, G., \& {Kitchatinov}, L.~L. 2011, \aap, 530,
  A48

\bibitem[{{Lagrange} {et~al.}(2010){Lagrange}, {Desort}, \&
  {Meunier}}]{lagrange10b}
{Lagrange}, A.-M., {Desort}, M., \& {Meunier}, N. 2010, \aap, 512, A38

\bibitem[{{Lagrange} {et~al.}(2011){Lagrange}, {Meunier}, {Desort}, \&
  {Malbet}}]{lagrange11}
{Lagrange}, A.-M., {Meunier}, N., {Desort}, M., \& {Malbet}, F. 2011, \aap,
  528, L9

\bibitem[{{Lanza} {et~al.}(2016){Lanza}, {Molaro}, {Monaco}, \&
  {Haywood}}]{lanza16}
{Lanza}, A.~F., {Molaro}, P., {Monaco}, L., \& {Haywood}, R.~D. 2016, \aap,
  587, A103

\bibitem[{{Lockwood} {et~al.}(2007){Lockwood}, {Skiff}, {Henry}, {Henry},
  {Radick}, {Baliunas}, {Donahue}, \& {Soon}}]{lockwood07}
{Lockwood}, G.~W., {Skiff}, B.~A., {Henry}, G.~W., {et~al.} 2007, \apjs, 171,
  260

\bibitem[{{Lovis} {et~al.}(2011){Lovis}, {Dumusque}, {Santos}, {Bouchy},
  {Mayor}, {Pepe}, {Queloz}, {S{\'e}gransan}, \& {Udry}}]{lovis11b}
{Lovis}, C., {Dumusque}, X., {Santos}, N.~C., {et~al.} 2011, ArXiv e-prints
  1107.5325

\bibitem[{{Mamajek} \& {Hillenbrand}(2008)}]{mamajek08}
{Mamajek}, E.~E. \& {Hillenbrand}, L.~A. 2008, \apj, 687, 1264

\bibitem[{{Martinez Pillet} {et~al.}(1993){Martinez Pillet}, {Moreno-Insertis},
  \& {Vazquez}}]{martinez93}
{Martinez Pillet}, V., {Moreno-Insertis}, F., \& {Vazquez}, M. 1993, \aap, 274,
  521

\bibitem[{{McQuillan} {et~al.}(2014){McQuillan}, {Mazeh}, \&
  {Aigrain}}]{mcquillan14}
{McQuillan}, A., {Mazeh}, T., \& {Aigrain}, S. 2014, \apjs, 211, 24

\bibitem[{{Messina} \& {Guinan}(2003)}]{messina03}
{Messina}, S. \& {Guinan}, E.~F. 2003, \aap, 409, 1017

\bibitem[{{Meunier}(2018)}]{meunier18a}
{Meunier}, N. 2018, \aap, 615, A87

\bibitem[{{Meunier} {et~al.}(2010{\natexlab{a}}){Meunier}, {Desort}, \&
  {Lagrange}}]{meunier10a}
{Meunier}, N., {Desort}, M., \& {Lagrange}, A.-M. 2010{\natexlab{a}}, \aap,
  512, A39

\bibitem[{Meunier \& Lagrange(2019{\natexlab{a}})}]{meunier18c}
Meunier, N. \& Lagrange, A.-M. 2019{\natexlab{a}}, submitted to A\&A

\bibitem[{Meunier \& Lagrange(2019{\natexlab{b}})}]{meunier18d}
Meunier, N. \& Lagrange, A.-M. 2019{\natexlab{b}}, submitted to A\&A

\bibitem[{{Meunier} {et~al.}(2015){Meunier}, {Lagrange}, {Borgniet}, \&
  {Rieutord}}]{meunier15}
{Meunier}, N., {Lagrange}, A.-M., {Borgniet}, S., \& {Rieutord}, M. 2015, \aap,
  583, A118

\bibitem[{{Meunier} {et~al.}(2010{\natexlab{b}}){Meunier}, {Lagrange}, \&
  {Desort}}]{meunier10}
{Meunier}, N., {Lagrange}, A.-M., \& {Desort}, M. 2010{\natexlab{b}}, \aap,
  519, A66

\bibitem[{{Meunier} {et~al.}(2017{\natexlab{a}}){Meunier}, {Lagrange}, {Mbemba
  Kabuiku}, {Alex}, {Mignon}, \& {Borgniet}}]{meunier17}
{Meunier}, N., {Lagrange}, A.-M., {Mbemba Kabuiku}, L., {et~al.}
  2017{\natexlab{a}}, \aap, 597, A52

\bibitem[{{Meunier} {et~al.}(2017{\natexlab{b}}){Meunier}, {Mignon}, \&
  {Lagrange}}]{meunier17b}
{Meunier}, N., {Mignon}, L., \& {Lagrange}, A.-M. 2017{\natexlab{b}}, \aap,
  607, A124

\bibitem[{{Mittag} {et~al.}(2013){Mittag}, {Schmitt}, \&
  {Schr{\"o}der}}]{mittag13}
{Mittag}, M., {Schmitt}, J.~H.~M.~M., \& {Schr{\"o}der}, K.-P. 2013, \aap, 549,
  A117

\bibitem[{{Morris} {et~al.}(2017){Morris}, {Hebb}, {Davenport}, {Rohn}, \&
  {Hawley}}]{morris17}
{Morris}, B.~M., {Hebb}, L., {Davenport}, J.~R.~A., {Rohn}, G., \& {Hawley},
  S.~L. 2017, \apj, 846, 99

\bibitem[{{Moss} {et~al.}(2011){Moss}, {Sokoloff}, \& {Lanza}}]{moss11}
{Moss}, D., {Sokoloff}, D., \& {Lanza}, A.~F. 2011, \aap, 531, A43

\bibitem[{{Nielsen} {et~al.}(2013){Nielsen}, {Gizon}, {Schunker}, \&
  {Karoff}}]{nielsen13}
{Nielsen}, M.~B., {Gizon}, L., {Schunker}, H., \& {Karoff}, C. 2013, \aap, 557,
  L10

\bibitem[{{Norris}(2018)}]{norris18}
{Norris}, C. 2018, PhD thesis, Imperial College London

\bibitem[{{Norris} {et~al.}(2016){Norris}, {Beeck}, {Unruh}, {Solanki}, {Yeo},
  \& {Krivova}}]{norris16}
{Norris}, C.~M., {Beeck}, B., {Unruh}, Y., {et~al.} 2016, in 19th Cambridge
  Workshop on Cool Stars, Stellar Systems, and the Sun (CS19), 63

\bibitem[{{Norris} {et~al.}(2017){Norris}, {Beeck}, {Unruh}, {Solanki},
  {Krivova}, \& {Yeo}}]{norris17}
{Norris}, C.~M., {Beeck}, B., {Unruh}, Y.~C., {et~al.} 2017, \aap, 605, A45

\bibitem[{{Noyes} {et~al.}(1984{\natexlab{a}}){Noyes}, {Hartmann}, {Baliunas},
  {Duncan}, \& {Vaughan}}]{noyes84}
{Noyes}, R.~W., {Hartmann}, L.~W., {Baliunas}, S.~L., {Duncan}, D.~K., \&
  {Vaughan}, A.~H. 1984{\natexlab{a}}, \apj, 279, 763

\bibitem[{{Noyes} {et~al.}(1984{\natexlab{b}}){Noyes}, {Weiss}, \&
  {Vaughan}}]{noyes84b}
{Noyes}, R.~W., {Weiss}, N.~O., \& {Vaughan}, A.~H. 1984{\natexlab{b}}, \apj,
  287, 769

\bibitem[{{Ol{\'a}h} {et~al.}(2016){Ol{\'a}h}, {K{\H o}v{\'a}ri}, {Petrovay},
  {Soon}, {Baliunas}, {Koll{\'a}th}, \& {Vida}}]{olah16}
{Ol{\'a}h}, K., {K{\H o}v{\'a}ri}, Z., {Petrovay}, K., {et~al.} 2016, \aap,
  590, A133

\bibitem[{{Ol{\'a}h} {et~al.}(2009){Ol{\'a}h}, {Koll{\'a}th}, {Granzer},
  {Strassmeier}, {Lanza}, {J{\"a}rvinen}, {Korhonen}, {Baliunas}, {Soon},
  {Messina}, \& {Cutispoto}}]{olah09}
{Ol{\'a}h}, K., {Koll{\'a}th}, Z., {Granzer}, T., {et~al.} 2009, \aap, 501, 703

\bibitem[{{Panja} {et~al.}(2019){Panja}, {Cameron}, \& {Solanki}}]{panja19}
{Panja}, M., {Cameron}, R., \& {Solanki}, S. 2019, in preparation

\bibitem[{{Radick} {et~al.}(2018){Radick}, {Lockwood}, {Henry}, {Hall}, \&
  {Pevtsov}}]{radick18}
{Radick}, R.~R., {Lockwood}, G.~W., {Henry}, G.~W., {Hall}, J.~C., \&
  {Pevtsov}, A.~A. 2018, \apj, 855, 75

\bibitem[{{Radick} {et~al.}(1998){Radick}, {Lockwood}, {Skiff}, \&
  {Baliunas}}]{radick98}
{Radick}, R.~R., {Lockwood}, G.~W., {Skiff}, B.~A., \& {Baliunas}, S.~L. 1998,
  \apjs, 118, 239

\bibitem[{{Reiners} {et~al.}(2016){Reiners}, {Mrotzek}, {Lemke}, {Hinrichs}, \&
  {Reinsch}}]{reiners16}
{Reiners}, A., {Mrotzek}, N., {Lemke}, U., {Hinrichs}, J., \& {Reinsch}, K.
  2016, \aap, 587, A65

\bibitem[{{Reiners} {et~al.}(2013){Reiners}, {Shulyak}, {Anglada-Escude},
  {Jeffers}, {Morin}, {Zechmeister}, {Kochukhov}, \& {Piskunov}}]{reiners13}
{Reiners}, A., {Shulyak}, D., {Anglada-Escude}, G., {et~al.} 2013, ArXiv
  e-prints 1301.2951

\bibitem[{{Reinhold} \& {Arlt}(2015)}]{reinhold15b}
{Reinhold}, T. \& {Arlt}, R. 2015, \aap, 576, A15

\bibitem[{{Reinhold} \& {Gizon}(2015)}]{reinhold15}
{Reinhold}, T. \& {Gizon}, L. 2015, \aap, 583, A65

\bibitem[{{Roudier} {et~al.}(2016){Roudier}, {Malherbe}, {Rieutord}, \&
  {Frank}}]{roudier16}
{Roudier}, T., {Malherbe}, J.~M., {Rieutord}, M., \& {Frank}, Z. 2016, \aap,
  590, A121

\bibitem[{{Saar} \& {Brandenburg}(1999)}]{saar99}
{Saar}, S.~H. \& {Brandenburg}, A. 1999, \apj, 524, 295

\bibitem[{{Saar} \& {Brandenburg}(2002)}]{saar02}
{Saar}, S.~H. \& {Brandenburg}, A. 2002, Astronomische Nachrichten, 323, 357

\bibitem[{{Saar} \& {Donahue}(1997)}]{saar97}
{Saar}, S.~H. \& {Donahue}, R.~A. 1997, \apj, 485, 319

\bibitem[{{Saar} {et~al.}(2003){Saar}, {Hatzes}, {Cochran}, \&
  {Paulson}}]{saar03}
{Saar}, S.~H., {Hatzes}, A., {Cochran}, W., \& {Paulson}, D. 2003, in Cambridge
  Workshop on Cool Stars, Stellar Systems, and the Sun, Vol.~12, The Future of
  Cool-Star Astrophysics: 12th Cambridge Workshop on Cool Stars, Stellar
  Systems, and the Sun, ed. A.~{Brown}, G.~M. {Harper}, \& T.~R. {Ayres},
  694--698

\bibitem[{{Samadi} {et~al.}(2007){Samadi}, {Georgobiani}, {Trampedach},
  {Goupil}, {Stein}, \& {Nordlund}}]{samadi07}
{Samadi}, R., {Georgobiani}, D., {Trampedach}, R., {et~al.} 2007, \aap, 463,
  297

\bibitem[{{Santos} {et~al.}(2015){Santos}, {Cunha}, {Avelino}, \&
  {Campante}}]{santos15}
{Santos}, A.~R.~G., {Cunha}, M.~S., {Avelino}, P.~P., \& {Campante}, T.~L.
  2015, \aap, 580, A62

\bibitem[{{Santos} {et~al.}(2017){Santos}, {Cunha}, {Avelino}, {Garc{\'{\i}}a},
  \& {Mathur}}]{santos17}
{Santos}, A.~R.~G., {Cunha}, M.~S., {Avelino}, P.~P., {Garc{\'{\i}}a}, R.~A.,
  \& {Mathur}, S. 2017, \aap, 599, A1

\bibitem[{{Scherrer} {et~al.}(1995){Scherrer}, {Bogart}, {Bush}, {Hoeksema},
  {Kosovichev}, {Schou}, {Rosenberg}, {Springer}, {Tarbell}, {Title},
  {Wolfson}, {Zayer}, \& {MDI Engineering Team}}]{Smdi95}
{Scherrer}, P.~H., {Bogart}, R.~S., {Bush}, R.~I., {et~al.} 1995, \solphys,
  162, 129

\bibitem[{{Schrijver}(2001)}]{schrijver01}
{Schrijver}, C.~J. 2001, Astrophys. J., 547, 475

\bibitem[{{Schr{\"o}der} {et~al.}(2012){Schr{\"o}der}, {Mittag}, {P{\'e}rez
  Mart{\'{\i}}nez}, {Cuntz}, \& {Schmitt}}]{schroder12}
{Schr{\"o}der}, K.-P., {Mittag}, M., {P{\'e}rez Mart{\'{\i}}nez}, M.~I.,
  {Cuntz}, M., \& {Schmitt}, J.~H.~M.~M. 2012, \aap, 540, A130

\bibitem[{{Schuessler} \& {Solanki}(1992)}]{schussler92}
{Schuessler}, M. \& {Solanki}, S.~K. 1992, \aap, 264, L13

\bibitem[{{Shapiro} {et~al.}(2014){Shapiro}, {Solanki}, {Krivova}, {Schmutz},
  {Ball}, {Knaack}, {Rozanov}, \& {Unruh}}]{shapiro14}
{Shapiro}, A.~I., {Solanki}, S.~K., {Krivova}, N.~A., {et~al.} 2014, \aap, 569,
  A38

\bibitem[{{Skumanich}(1972)}]{skumanich72}
{Skumanich}, A. 1972, \apj, 171, 565

\bibitem[{{Snodgrass} \& {Ulrich}(1990)}]{snodgrass90}
{Snodgrass}, H.~B. \& {Ulrich}, R.~K. 1990, \apj, 351, 309

\bibitem[{{Steiner} {et~al.}(2014){Steiner}, {Salhab}, {Freytag}, {Rajaguru},
  {Schaffenberger}, \& {Steffen}}]{steiner14}
{Steiner}, O., {Salhab}, R., {Freytag}, B., {et~al.} 2014, \pasj, 66, S5

\bibitem[{{Strassmeier} {et~al.}(2012){Strassmeier}, {Weber}, {Granzer}, \&
  {J{\"a}rvinen}}]{strassmeier12}
{Strassmeier}, K.~G., {Weber}, M., {Granzer}, T., \& {J{\"a}rvinen}, S. 2012,
  Astronomische Nachrichten, 333, 663

\bibitem[{{Su{\'a}rez Mascare{\~n}o} {et~al.}(2016){Su{\'a}rez Mascare{\~n}o},
  {Rebolo}, \& {Gonz{\'a}lez Hern{\'a}ndez}}]{suarez16}
{Su{\'a}rez Mascare{\~n}o}, A., {Rebolo}, R., \& {Gonz{\'a}lez Hern{\'a}ndez},
  J.~I. 2016, \aap, 595, A12

\bibitem[{{Unruh} {et~al.}(1999){Unruh}, {Solanki}, \& {Fligge}}]{unruh99}
{Unruh}, Y.~C., {Solanki}, S.~K., \& {Fligge}, M. 1999, \aap, 345, 635

\bibitem[{{Vidotto} {et~al.}(2014){Vidotto}, {Gregory}, {Jardine}, {Donati},
  {Petit}, {Morin}, {Folsom}, {Bouvier}, {Cameron}, {Hussain}, {Marsden},
  {Waite}, {Fares}, {Jeffers}, \& {do Nascimento}}]{vidotto14}
{Vidotto}, A.~A., {Gregory}, S.~G., {Jardine}, M., {et~al.} 2014, \mnras, 441,
  2361

\bibitem[{{Wilson}(1963)}]{wilson63}
{Wilson}, O.~C. 1963, \apj, 138, 832

\bibitem[{{Wright}(2005)}]{wright05}
{Wright}, J.~T. 2005, \pasp, 117, 657

\bibitem[{{Wright} {et~al.}(2004){Wright}, {Marcy}, {Butler}, \&
  {Vogt}}]{wright04}
{Wright}, J.~T., {Marcy}, G.~W., {Butler}, R.~P., \& {Vogt}, S.~S. 2004, \apjs,
  152, 261

\bibitem[{{Yeo} {et~al.}(2013){Yeo}, {Solanki}, \& {Krivova}}]{yeo13}
{Yeo}, K.~L., {Solanki}, S.~K., \& {Krivova}, N.~A. 2013, \aap, 550, A95

\end{thebibliography}

\begin{appendix}

\section{Rotation period}

\begin{figure}
\includegraphics{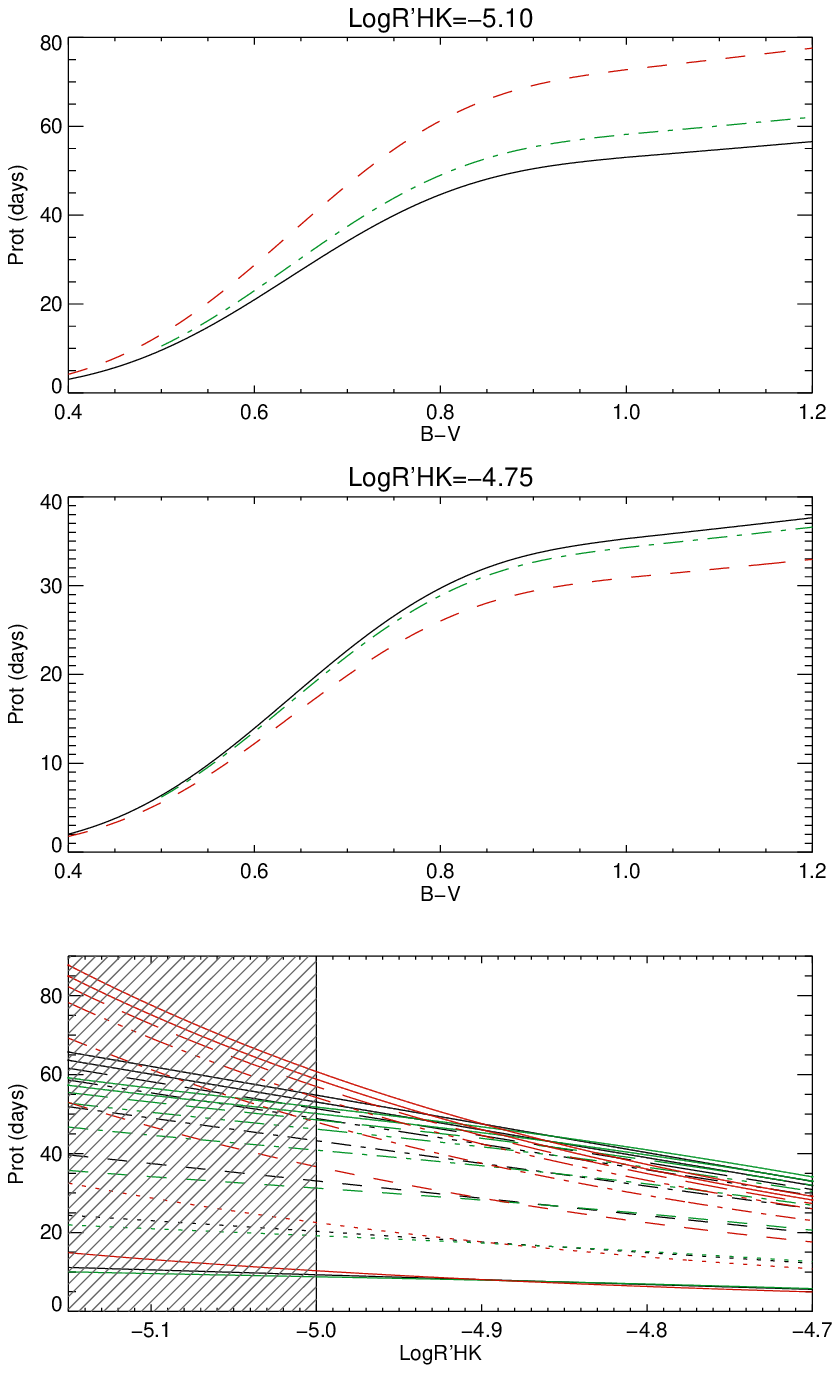}
\caption{
{\it First panel}: Comparison of the rotation period vs. B-V for LogR'$_{\rm HK}$ =-5.1 from various sources: the solid black line is from \cite{mamajek08}, the green dot-dashed line from \cite{noyes84}, and the dashed red line from \cite{saar99}. 
{\it Second panel}: Same for LogR'$_{\rm HK}$=-4.75. 
{\it Third panel}: Similar comparison vs. LogR'$_{\rm HK}$ for eight levels in B-V between 0.5 and 1.1. The vertical line approximately shows the lower limit for the validity of the laws (the shaded area indicates the zones where it is not valid).
}
\label{prot_annexe}
\end{figure}

Fig.~\ref{prot_annexe} shows a comparison of the rotation period derived from observations presented in different sources. The first two plots show the rotation period versus B-V for two values of LogR'$_{\rm HK}$ (-5.10 and -4.75, respectively) from \cite{mamajek08}, \cite{noyes84}, and \cite{saar99}.
\cite{mamajek08} and \cite{noyes84} are quite close to each other. \cite{saar99} give rotation periods longer than all the others for the most quiet stars, as illustrated in the last plot, but these are always poorly constrained. 



Other papers also provide rotation periods for very large samples, for example, from Kepler light curves \cite[][]{nielsen13,mcquillan14,garcia14,reinhold15}, but they depend on the photometric variability, which cannot be translated directly into an average LogR'$_{\rm HK}$ level. 
Others are only given as a function of magnetic fields \cite[][]{vidotto14} or without any indication of activity level \cite[][]{strassmeier12}.
They can therefore not be easily included in our set of parameters as such.

\section{Differential rotation and maximum latitude}

\cite{reinhold15} compared their measured differential rotations with previous results. They obtained a good agreement with laws obtained by \cite{hall91} and \cite{donahue96}. \cite{barnes05} and \cite{colliercameron07} obtained much steeper laws versus T$_{\rm eff}$ , however.  We note that \cite{daschagas16} obtained a differential rotation that is very similar to the solar one from the analysis of 17 Kepler solar-type stars. The results obtained by \cite{balona16} are more difficult to compare because of their normalization. 
Most numerical simulations of stellar differential rotation cover a small range in parameters, either have very fast rotation periods \cite[e.g.,][]{kuker11}, or they are too close to the solar case \cite[e.g.,][]{kuker11b}. The simulations of \cite{brun17} cover a wider range (G and K stars): They derive a scaling law of $\Delta \Omega$ versus mass and $\Omega$. For their solar-type differential rotation, they obtained a scaling as M$^{0.73}$$\Omega^{0.66}$, for  $\Delta \Omega$ between equator and 60$^{\circ}$. They attempted a comparison with the scaling laws in the literature, which they found to be not very conclusive, but they did not discuss the effect of $\theta_{\rm max}$ on the observations. It is therefore difficult to use these laws to build our simulations.

\section{Computation of the observables}

We detail here how the observables were computed at each time step to produce the time series. The sum in all formulae is made on all structures of a given type at the corresponding time step, and all observables below are functions of time. 
We use the following notations: ${\rm A}_j$ is the size of the structure j (in ppm of the hemisphere); $\theta_j$ and $\phi_j$ are their latitude (between -90$^{\circ}$ and 90$^{\circ}$) and longitude (between 0$^{\circ}$ and 360$^{\circ}$), respectively; $\mu_j$ is their position on the disk (cosine between the local surface and the line of sight, it takes a value of 1 at disk center and zero at the limb, and depends on $\theta_j$, $\phi_j$, and inclination).

Other variables are as follows: $i$ is the star inclination; ${\rm Pcb}$ is the center-to-limb darkening function at a given temperature, as in \cite{borgniet15}, from \cite{claret03}, using a log(g) of 4.5 and solar metallicity; ${\rm C_{pl}}$ is the relative contrast of the plages and is a function of $\mu$; subscripts "phot", "pl", and "sp" are for photosphere, plages (including network), and spots, respectively; and T$_{\rm eff}$ is the photospheric temperature.  

\subsection{Filling factors of the spots and plages}

The filling factor of either spots or plages is defined as

\begin{equation}
{\rm ff}=\Sigma A_j \times \mu_j \times 2
\end{equation}

and is in ppm of the stellar disk. 


\subsection{Photometry of the spots and plages}

The plage contribution to the photometry is defined as

\begin{multline}
{\rm I_{pl}} = \Sigma {\rm A}_j \times \mu_j \times {\rm C_{pl}}(\mu_j,{\rm A}_j,{\rm T_{eff}}) \times 2 \\
\times {\rm f_{phot}}({\rm T_{eff}}) \times {\rm Pcb}(\mu_j,{\rm T_{eff}}) / \Phi_{\rm tot}({\rm T_{eff}}),
\end{multline}



where ${\rm C_{pl}}$ is a relative contrast DeltaI/I=(Ipl-Iphot)/Iphot (local (=Iplage/Iphot -1))
and 
$\Phi_{\rm tot}$ is equal to ${\rm f_{phot}}$  $\times$ Pcb then integrated over the disk, where f is the Planck function for the indicated temperature. 

The spot contribution to the photometry is defined as

\begin{equation}
{\rm I_{sp}} = \Sigma {\rm A}_j \times \mu_j \times {\rm C_{sp}}(\mu_j,{\rm T_{sp}},{\rm T_{eff}}) \times 2 
,\end{equation}


where 

\begin{multline}
{\rm C_{sp}} = ({\rm f_{sp}}({\rm T_{sp}}) \times {\rm Pcb}(\mu_j,{\rm T_{sp}}) - {\rm f_{ph}}({\rm T_{eff}}) \times {\rm Pcb}(\mu_j,{\rm T_{eff}})) \\
/ \Phi_{\rm tot}({\rm T_{eff}}).
\end{multline}

f and $\Phi_{\rm tot}$ are defined as above. 
The photometric contributions are in ppm of the quiet-star brightness. The sum of the two provides the total variability in ppm.
 
\subsection{RV of spots, plages, and convection inhibition}

The plage contribution to the RV is defined as

\begin{multline}
{\rm RV_{pl}} = \Sigma {\rm A}_j \times \mu_j \times 2 10^{-6} \times \Omega(\theta_j) \times \sin(\phi_j) \times \sin(i) \times \\
{\rm C_{pl}}(\mu_j,{\rm A}_j,{\rm T_{eff}}) \times {\rm Pcb}(\mu_j,{\rm T_{eff}}) \times {\rm f_{ph}}({\rm T_{eff}}) / \Phi_{\rm tot}({\rm T_{eff}}),
\end{multline}

where $\Omega$ is the rotation rate converted in m/s. 

The inhibition of the convection contribution to the RV is defined as

\begin{multline}
{\rm RV_{conv}} = \Sigma {\rm A}_j \times \mu_j \times 2 10^{-6} \times {\rm \Delta V}(\mu_j) \times {\rm Pcb}(\mu_j,{\rm T_{eff}}) \times \\
(1+ {\rm C_{pl}}(\mu_j,{\rm A}_j,{\rm T_{eff}}) ) \times {\rm f_{ph}}({\rm T_{eff}}) / \Phi_{\rm tot}({\rm T_{eff}}),
\end{multline}

where $\Delta V$ is the attenuation of the convective blueshift (vertical contribution, see Sect.~3.2.1 for a discussion).

The spot contribution to the RV is defined as

\begin{multline}
{\rm RV_{sp}} = \Sigma {\rm A}_j \times \mu_j \times 2 10^{-6} \times \Omega(\theta_j) \times \sin(\phi_j) \times \sin(i) \\
\times {\rm C_{sp}}(\mu_j,{\rm T_{sp}},{\rm T_{eff}}), 
\end{multline}

where ${\rm C_{sp}}$ is defined as above. 
All RV are in m/s. The Zeeman effect \cite[][]{reiners13} is not taken into account. 
The sum of the three components then provides the total RV variation in m/s.

\subsection{Astrometric displacements}

The astrometric contribution in the x direction (x taken along the equator) is the same for plages and spots: 
\begin{multline}
{\rm \Delta x} = \Sigma {\rm I}_j \times {\rm x}_j \times R(R_\odot) \times 180 \times 3600 \times / {\rm D_{star}}/\pi \times {\rm N,}
\end{multline}


where ${\rm I}_j$ is the individual contribution of a structure to the photometry (see Eq C.2 and C.3). It is normalized to the quiet star. The formula is similar in y (along the rotation axis): ${\rm x}_j$ and ${\rm y}_j$ are positions in this referential system, relative to the stellar radius. 
${\rm D_{star}}$ is arbitrarily chosen to be 10 pc (for a star at a different distance, the corresponding factor must be applied).
$N$ is a normalization factor equal to  1/(1+${\rm I_{pl}}$+${\rm I_{sp}}$) to normalize with the actual flux of the star.
Astrometric displacements are in $\mu$as. 

\section{Chromospheric emission model: practical recipe}

\subsection{Model for a G2 star}

We followed the approach described in \cite{meunier18a}, with some simplifications because magnetic field maps are not available. The S-index model consists of three components and is described as

\begin{equation}
S(t)=S_{\rm basal}+[S_{\rm act}(t)+ S_{\rm qs}(t)]  \times f / N
,\end{equation}

where $N$ is equal here to 10$^6$ (areas are in ppm) and the three contributions are therefore the following : 
\begin{itemize}

\item{{\it \textup{The basal flux}} corresponds to stars with no activity \cite[see discussion in][]{meunier18a}, and must not be confused with the lower limit in LogR'$_{\rm HK}$ discussed in Sect.~2.3.1. For B-V lower than 0.94, we used the basal S-index of 0.144 derived by \cite{mittag13}. For B-V above 0.94, the true basal flux is significantly below the lower limit observed in LogR'$_{\rm HK}$:  We used the basal flux built from models and observations made by \cite{schroder12}. The corresponding basal flux used in this paper is shown as the lower solid line in Fig.~\ref{bv_logrphk}. }
\item{{\it \textup{The active component}} is due to the plages and network features that are simulated in Sect.~2. We attributed a magnetic field to each size. The sizes used in this work correspond to a threshold of 100~G on MDI/SOHO \cite{Smdi95} magnetograms (see Sect.~2.1.1), as defined in \cite{meunier10a}. These sizes were then converted into sizes corresponding to a 40~G threshold as in \cite{meunier18a} before we used the scaling law of the magnetic field versus size from MDI/SOHO.
Then we computed the contribution of each plage as a multiplying factor times this magnetic field to a certain exponent. The parameters were kept to the solar values (see Section D3 for a discussion) and depend on the size of the region, as detailed in \cite{meunier18a}.     
}
\item{
{\it \textup{The quiet-star component}} (QS) is due to the weak magnetic field everywhere else on the surface. This contribution is critical and necessary \cite[see][for a discussion]{meunier18a}: if we had kept the solar relationship between filling factor ff and S-index as obtained by \cite{shapiro14} for example, it would be impossible to model the S-index for stars with an activity level much below the minimum solar cycle (since ff=0 then). 
We assumed that the magnetic flux in the quiet star is related to the average activity level of the star: if it is more active, it provides more flux to the quiet star and therefore a higher chromospheric emission.  The procedure is detailed below.
}
\end{itemize}



We tested a description of the QS component similar to the one used in our solar model \cite[][]{meunier18a}, with the form f'$\times$B$_{\rm QS}^{\beta}$, where B$_{\rm QS}$ is the average magnetic field in the quiet Sun. It is difficult to apply directly, however: If f' is kept to the solar value, this contribution remains strong even for very low values of B$_{\rm QS}$. If we adapt f' to B$_{\rm QS}$ by allowing it to decrease sharply toward zero at low activity levels, it leads to very strong variations of the S-index with time due to the quiet star, which does not seems realistic. A QS contribution that is too large for quiet stars prevents us from modeling stars with a large cycle amplitude because the expected level at cycle minimum would then fall below the basal+QS level.
Instead, we chose an empirical approach, given the lack of constraints, which allowed us to model stars with a large amplitude as observed. For a given cycle amplitude in our (B-V,LogR'$_{\rm HK}$) grid, given the average LogR'$_{\rm HK}$ and the minimum LogR'$_{\rm HK}$ corresponding to the amplitude, it is possible to compute the maximum QS contribution that we should allow to model such variability. Then we chose a QS level slighly below the minimum value: this imposed a small number of spots at cycle minimum, typically about 2-4.
A more sophisticated model would require numerical simulations of magnetic flux tubes from active regions down to very small scales and is beyond the scope of this paper.

\subsection{S-index for other stars}


This S-index, built as in Sect.~D.1, is valid for G2 stars (because it has been validated for the Sun) but not for other stars: the pertinent value to compare chromospheric emissions between stars is the LogR'$_{\rm HK}$ , however. 
For a given simulation at B-V, we transformed the obtained S-index into a LogR'$_{\rm HK}$ using the solar value (0.65), then back into an S-index with the stellar LogR'$_{\rm HK}$. In practice, we mostly used the LogR'$_{\rm HK}$.  

\subsection{Possible intrisinc chromospheric variability versus B-V?}

A dependence of the chromospheric emission (for similar plages) on the spectral type is not taken into account in the previous model. There have been very few studies of the chromospheric emission that would allow us to answer this question precisely. Numerical simulations of magnetic structures \cite[][]{steiner14,beeck15} suggest that the magnetic field remains similar in their magnetic regions (and if the magnetic flux increases, it is mostly due to the size, not the flux density). Therefore, we do not expect large variations. \cite{cuntz99} expected the rotation period to affect the chromospheric emission because a low rotation would lead to a smaller spreading of the flux tubes, which in turn would cause a slightly higher chromosperic emission, but it is difficult to directly apply to our parameters. \cite{fawzy02b}, based on models of magnetic waves propagating in flux tubes to explain the chromospheric heating \cite[][]{fawzy02a,fawzy02c}, established a relation between chromospheric emission and magnetic coverage. They obtained an increasing emission for larger B-V. We can derive an average trend from their results, which would give a corrective factor equal to 1.12-0.18 $\times$(B-V). However, this is very uncertain because they specified that they might have uncertainties of up to a factor two on the flux emissions.  Therefore, we chose not to include this trend in our model. 

\section{Input parameters}

\begin{figure*}
\includegraphics{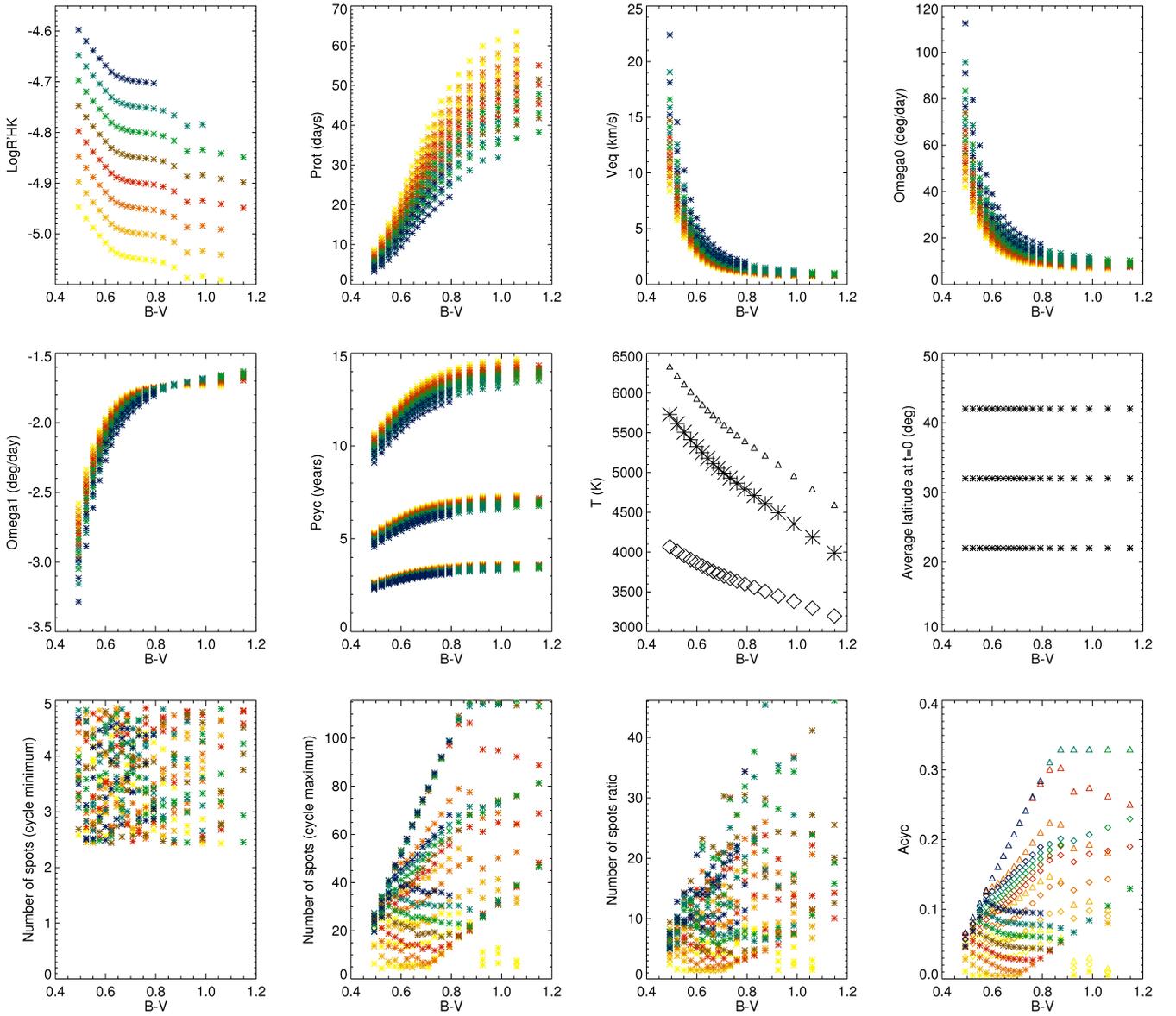}
\caption{
Summary of the input parameters vs. B-V, illustrating the range they cover for each spectral type, for the 11421 simulations. All activity levels are superimposed. The color code corresponds to different average activity levels (well illustrated in the first panel). Temperatures correspond to T$_{\rm eff}$ (triangles), solar spot contrast (stars), and the largest spot contrast (diamonds).  
}
\label{recap_param}
\end{figure*}

In this appendix we summarize the parameters. 
Table~\ref{tab_param} shows the list of parameters and their values (or ranges). These were used to produce the list of spots, plages, and network features at each time step (size and position). 
Table~\ref{tab_param2} shows the same summary for the parameters that allowed us to compute the observables. 
Finally, Fig.~\ref{recap_param} illustrates the range of values covered by the parameters, which are adapted to each spectral type as a function of B-V.

\begin{table*}
\caption{Input parameters to generate spots and plages}
\label{tab_param}
\begin{center}
\renewcommand{\footnoterule}{}  
\begin{tabular}{lllll}
\hline
Category & Parameter  & Value/Range &   Unit &  Reference \\ 
\hline
Solar activity &        Cycle shape &   Smoothed average & -  & Sect.~3.6.3\\
 &       &  solar cycle &   & \\
  & Cycle period &      2-14.5 &        [year]  & Sect.~3.6.1 \cite{noyes84b}\\
                &  &  &  & \cite{baliunas96} \\
                &  &  &  & \cite{saar99}\\
                &  &  &  & \cite{bohm07}\\
                &  &  &  & \cite{olah09,lovis11b}\\
                &  &  &  & \cite{suarez16}\\
                &  &  &  & \cite{olah16}\\
  & Cycle amplitude &   0.03-0.43 &     [LogR'$_{\rm HK}$]      & Sect.~3.6.2 \cite{lovis11b}\\
                &  &  &  & \cite{radick98} \\
 &       spot number random dispersion &        25      & \% &   Adapted from \cite{borgniet15} \\       
\hline
Stellar & &  &  &  \\
fondamental &  Stellar radius &   0.9-1.4 &  [R$_{\odot}$] & Sect.~3.2 \cite{boyajian12,boyajian13} \\ 
parameters   &   Stellar T$_{\rm eff}$ & 4594-6334 & [K] & Sect.~3.2 \cite{gray05} \\ 
\hline
Spatio-temporal &       Mean start latitude  &   22, 32, 42 &   [deg]  &  Sect.~3.5 \\   
distribution &  {\it Mean end latitude} &        {\it 9} &      [deg]    & \cite{borgniet15}  \\
&       {\it Standard lat. dispersion} &        {\it 6} &       [deg]    &  id. \\
&       {\it Max. lat. dispersion} &    {\it 20} &      [deg]    & id.  \\
&       {\it North-south asymmetry} &   {\it 0.5}& -  &  id. \\
&       {\it Active longitude spot fraction} &  {\it 0.4}& -     & id.  \\
&       {\it Active longitude extension area}&   {\it +/-20}&   [deg]    & id.  \\
\hline
Large scale &   Differential rotation&  $\Omega_0$ = 6.6-112.5 &        [deg/day] & Sect. 3.5 adapted from \\
dynamics                &  &  &  & \cite{mamajek08}\\
                &  &  &  & \cite{reinhold15}\\
       & &            $\Omega_1$ = -2.10 -- -7.25 &   [deg/day]       & id.\\
&       {\it Meridional flow} & $\alpha$ = {\it 12.9} &  [m/s] &        \cite{komm93}\\
 & &            $\beta$ = {\it 1.4} &   [m/s] & id. \\
\hline
Isolated &      {\it Total fraction}&   {\it 0.4} &     - &     \cite{martinez93}\\
spots & {\it Mean initial size}&        {\it 46.51}&    [$\mu$Hem] & adapted from    \cite{baumann05} \\
properties&     {\it Standard size deviation}&  {\it 2.14} &    [$\mu$Hem] &       id. \\
&       {\it Max. size} &       {\it 1500 }&    [$\mu$Hem]      &Papers I and II, \cite{borgniet15}\\
&       {\it Mean decay}&       {\it -18.9}&    [$\mu$Hem/day]&         \cite{martinez93}\\
&       {\it Median decay}&     {\it -14.8}&    [$\mu$Hem/day]&         id. \\
\hline
Complex &       {\it Total fraction} &  {\it 0.6} &     -& \cite{martinez93}\\
spots & {\it Mean initial size}&        {\it 90.24} &   [$\mu$Hem]& adapted from \cite{baumann05}\\
properties &       {\it Standard size deviation}&  {\it 2.49} &    [$\mu$Hem]&     id. \\
&       {\it Max. size}&        {\it 5000} &    [$\mu$Hem] &    Papers I and II, \cite{borgniet15}\\
&       {\it Mean decay}&       {\it -41.3}&    [$\mu$Hem/day]&\cite{martinez93}\\
&       {\it Median decay} &    {\it -30.9}&    [$\mu$Hem/day]&         id. \\
\hline
All spots &     {\it Min. decay value}&         {\it -3} &      [$\mu$Hem/day]          &  id. \\
&       {\it Max. decay value}&         {\it -200} &    [$\mu$Hem/day]   & id.  \\
&       {\it Min. spot size} &  {\it 10} &      [$\mu$Hem] &    Papers I and II, \cite{borgniet15}\\
\hline
Faculae  &      {\it q (facula-to-spot ratio)} &  & &  \cite{borgniet15}\\                      
properties &    {\it Mean log(q)}&      {\it 0.8}&      - &      id.  \\
&       {\it Standard deviation (log(q))} &     {\it 0.4} &     -       & id.  \\
&       {\it Min.- Max. log(q)} &       {\it 0.1--5} &  -       &   id. \\
&       {\it Mean decay}& {\it  -27} &  [$\mu$Hem/day]  &  id.  \\
&       {\it Median decay}& {\it        -20} &  [$\mu$Hem/day]  &  id.  \\
&       {\it Min. facula size}& {\it    3}      &[$\mu$Hem] &   Papers I and II\\
\hline
Network &       Diffusion coefficient&  69-407 & [km$^2$/s] & Sect.~3.7.2 \cite{schrijver01}\\
properties                & & & & \cite{meunier17b}\\
            &    {\it Remainder fraction for decay}&     {\it 0.975}&    [ /day]   & \cite{borgniet15}   \\
&       {\it Min. size}&        {\it 3} &       [$\mu$Hem] &    Papers I and II, \cite{borgniet15}\\
&       {\it Facula fraction recovered} &       {\it 0.8} &     -  & \cite{borgniet15}  \\     
\hline
\end{tabular}
\end{center}
\tablefoot{
Values kept to the solar values used in \cite{borgniet15} are in italics.
}
\end{table*}

\begin{table*}
\caption{Input parameters to generate the time series (photometry, astrometry, and radial velocity)}
\label{tab_param2}
\begin{center}
\renewcommand{\footnoterule}{}  
\begin{tabular}{lllll}
\hline
Category & Parameter  & Value & Unit &  Reference \\ 
\hline
 &      Spot T contrast  & 600-2600 & [K]  & Sect.~3.8 \cite{berd05,borgniet15}\\
 &  Plage contrast & 0.02-0.13  & -  & adapted from \cite{unruh99,meunier10a}\\
   &  &  &  &  \cite{borgniet15,norris18}  \\
 &  Convective blueshift $\Delta$V & 90-520  & [m/s]  & Sect.~3.7.1 \cite{meunier17b}  \\
 &  Attenuation factor of $\Delta$V &  0.37 &   &  Sect.~3.7.1\cite{meunier17b}  \\
 &  {\it Limb-darkening coefficients} & {\it f(T$_{\rm eff}$)}  & -  & \cite{claret03}  \\
 &  Chromospheric emission parameters &   &   & Sect.~4 \cite{harvey99,meunier18a}  \\
\hline
\end{tabular}
\end{center}
\tablefoot{
Values kept to the solar values or laws used in \cite{borgniet15} are in italics.
}
\end{table*}


\begin{table*}
\caption{Summary of the laws corresponding to Fig.~1}
\label{tab_law}
\begin{center}
\renewcommand{\footnoterule}{}  
\begin{tabular}{lll}
\hline
Variable & Law  & Section \\ 
\hline
Lower limit in activity level & S=0.144 for B-V$<$0.94 & 2.3.1 \\
  & S= 0.0269231$\times$(B-V)+0.118892 for B-V$>$0.94 &        \\
Upper limit in activity level & LogR'$_{\rm HK}$=-0.375$\times$(B-V)-4.4 &  2.3.2 \\
P$_{\rm rot}$ (days) & P$_{\rm rot}$=  (R$_0$+$\delta$)$\times \tau_c$    &  2.4 \\
   & with R$_0$=0.808-2.966$\times$(LogR'$_{\rm HK}$+4.52) &  \\
   & and $\delta$=$[$-0.2,0,0.2$]$ & \\
Differential rotation   & Log($\alpha$)=$p_0$(T$_{\rm eff}$)+$p_1$(T$_{\rm eff}$)$\times$Log(P$_{\rm rot}$) (*) &  2.5 \\
P$_{\rm cyc}$ (days)    & P$_{\rm cyc}$=(P$_{\rm rot}$$\times$10$^y$) & 2.6.1 \\
   & with y=0.84$\times$Log(1/P$_{\rm rot}$)+3.14+$\delta$ &   \\
   & and $\delta$=$[$-0.3,0,0.3$]$ & \\
Acyc (limits from Fig.~\ref{acyclovis}) & 0.727$\times$(B-V)-0.292 if B-V$<$0.851, 0.33 otherwise    & 2.6.7 \\
   & 0.28$\times$(B-V)-0.196    &  \\
   & 0.342$\times$LogR'$_{\rm HK}$+1.703    &  \\
Attenuation of the convective & $\Delta$V=-0.1718$\times$($p_0$(T$_{\rm eff}$)+$p_1$(T$_{\rm eff}$)$\times$LogR'$_{\rm HK,basal})$ & 2.7 \\
blueshift (m/s)    &   with p$_0$=198557.04-118.86301$\times$T$_{\rm eff}$+0.023348413$\times$T$_{\rm eff}$$^2$-1.4980577e-06$\times$T$_{\rm eff}$$^3$ & \\
   &  and p$_1$=46335.574-27.529228$\times$T$_{\rm eff}$+0.0053657815$\times$T$_{\rm eff}$$^2$-3.4108808e-07$\times$T$_{\rm eff}$$^3$ & \\
$\Delta$T$_{\rm spot}$ (upper limit, K)  & 0.75$\times$T$_{\rm eff}$-2250  &  2.8 \\
$\Delta$T$_{\rm spot}$ (lower limit, K)  &  605     &  2.8 \\
Plage contrast & C($\mu$,500G,T$_{\rm eff}$)=$\sum\limits_{i=0}^{4} c_i$(500G,T$_{\rm eff}) \times \mu^i $ (**) & 2.9 \\
   & C($\mu$,100G,T$_{\rm eff}$)=$\sum\limits_{i=0}^{4} c_i$(100G,T$_{\rm eff}) \times \mu^i $  & \\
\hline
\end{tabular}
\end{center}
\tablefoot{ 
(*) See Sect.~2.5 for more details. (**) The coefficients $c_i(500G,$T$_{\rm eff})$ and $c_i(100G,$T$_{\rm eff})$ have been derived from \cite{norris18} for M0, K0, and G2 stars and were then interpolated or extrapolated for each of our spectral types (see Sect.~2.9 for more details).  
}
\end{table*}


\end{appendix}

\end{document}